\documentclass[aps,pra,twocolumn,superscriptaddress,longbibliography]{revtex4-1}

\usepackage{latexsym}		
\usepackage{amsmath}
\usepackage{graphics}
\usepackage[usenames]{color}
\usepackage{color}
\usepackage{tikz}
\usepackage{amssymb}

\usepackage{braket}

\begin{document}
\newcommand{\pd}[2]{\frac{\partial #1}{\partial #2}}
\newcommand{\der}[2]{\frac{d #1}{d #2}}
\newcommand{\pdd}[2]{\frac{\partial^2 #1}{\partial #2^2}}
\newcommand{\secder}[2]{\frac{d^2 #1}{d #2^2}}
\newcommand{\NMcomment}[1]{{\bf \textcolor{red}{ #1}}}
\newcommand{\NMadd}[1]{{\color{red}{ #1}}}
\newcommand{\KHcomment}[1]{{\bf \textcolor{blue}{ #1}}}
\newcommand{\CTcomment}[1]{{\bf \textcolor{green}{ #1}}}

\newcommand{\warn}[1]{{\color{red}\textbf{ *  #1  *}}}
\newcommand{\warncite}[1]{{\color{blue}{ cite:  #1}}}

\title{A Model for Scattering with Proliferating Resonances: Many Coupled Square Wells}
\author{Nirav P. Mehta}
\email{nmehta@trinity.edu}
\affiliation{Trinity University, One Trinity Place, San Antonio, TX. 78212-7200 USA}
\affiliation{Kavli Institute of Theoretical Physics, University of California,
Santa Barbara, Santa Barbara, California 93106, USA}
\author{Kaden R.~A. Hazzard} \email{kaden.hazzard@gmail.com}
\affiliation{Department of Physics and Astronomy, Rice University, Houston, Texas 77005, USA}
\affiliation{Rice Center for Quantum Materials, Rice University, Houston, Texas 77005, USA}
\author{Christopher Ticknor}
\email{cticknor@lanl.gov}
\affiliation{Theoretical Division, Los Alamos National Laboratory, Los Alamos, New Mexico 87545, USA}
\affiliation{Kavli Institute of Theoretical Physics, University of California,
Santa Barbara, Santa Barbara, California 93106, USA}
\date{\today}

\begin{abstract}
   We present a multichannel model for elastic interactions, comprised of an arbitrary number of coupled finite square-well potentials, and derive  semi-analytic solutions for its scattering behavior. Despite the model's simplicity, it is flexible enough to include many coupled short-ranged resonances in the vicinity of the collision threshold, as is necessary to describe ongoing experiments in ultracold molecules and lanthanide atoms.    We also introduce a simple, but physically realistic, statistical ensemble for parameters in this model. We compute the resulting probability distributions of nearest-neighbor resonance spacings and analyze them by fitting to the Brody distribution. We quantify the ability of alternative distribution functions, for resonance spacing and  resonance number variance, to describe the crossover regime. The  analysis demonstrates that the multichannel square-well model  with the chosen ensemble of parameters naturally captures  the crossover from integrable to chaotic scattering as a function of closed channel coupling strength. 

\end{abstract}

\maketitle
 
\section{Introduction}

Recent experiments on ultracold ground state molecules  and lanthanide atoms  are expanding the purview of ultracold matter. Diverse  species are now  studied experimentally, including  homonuclear~\cite{danzl:ultracold_2010,reinaudi:optical_2012,stellmer:creation_2012,frisch:ultracold_2015}
and dipolar~\cite{ni:high_2008,		
park:ultracold_2015,
seesselberg:modeling_2018,takekoshi:ultracold_2014,molony:creation_2014,guo:creation_2016,rvachov:long_2017,demarco:fermi_2018}
molecules that are formed by cooling atoms and then coherently associating them, molecules that are laser-cooled~\cite{hummon:2D_2013,
barry:magneto_2014,
anderegg:radio_2017,
truppe:molecules_2017,
anderegg:laser_2018}, and a variety of lanthanide atoms~\cite{lu:strongly_2011,lu:quantum_2012,aikawa:bose-einstein_2012,aikawa:reaching_2014,trautmann:dipolar_2018,
sukachecv:magneto_2010,miao:magneto_2014}. These ultracold species furnish new capabilities for quantum computing~\cite{ni:dipolar_2018,
demille:quantum_2002,
andre:coherent_2006,
yelin:schemes_2006,
herrera:infrared_2014,
karra:prospects_2016}, exploring many-body phenomena~
\cite{wall:quantum_2015,	hazzard:many-body_2014,
yan:realizing_2013, carr:cold_2009,	baranov:condensed_2012,			bohn:cold_2017,lemeshko:manipulation_2013},
  measuring fundamental phenomena~\cite{kozlov:parity_1995,
flambaum:enhanced_2007,
hudson:improved_2011,
baron:order_2014,
cairncross:precision_2017}, and studying and controlling chemistry~\cite{ospelkaus:quantum_2010,	ni:dipolar_2010,miranda:controlling_2011,
balakrishnan:perspective_2016,krems:molecules:2005,krems:cold_2008,
liu:building_2018}.   
Many of these applications rely on the variety of internal states offered by these systems.
Lanthanide atoms possess numerous hyperfine states  and  electronic states resulting from their open $f$-shell. Molecules display even more internal states, arising from rotational and vibrational degrees of freedom. 
Although the numerous internal states lend new capabilities to ultracold matter, they also have dramatic effects on the interactions.

Consequently, unlocking  the potential of ultracold molecules and lanthanide atoms   requires us to dispense with a fundamental assumption about ultracold interactions. Specifically, it is no longer sufficient to approximate  interactions by commonly used effective potentials, such as a delta function pseudopotential $V(\vec{r})=g\delta(\vec{r})$ or a two-channel resonance model~\cite{Chin:2010rmp}.

To understand why such simple effective interactions fail to describe collisionally complex systems, let's recall how these interactions arise in more typical ultracold matter, for example, alkali or alkaline-earth atoms. The key idea  in the latter cases is that the collisions at ultracold temperatures are near in energy to at most one closed-channel bound state. Together with the short-ranged nature of the interactions, this implies that the scattering is energy-independent over the range of energies of interest in experiment and well-described by a simple effective interaction, such as the delta function pseudopotential. 
In addition to justifying simple effective interaction potentials, these conditions allow one to quantitatively predict how  magnetic and electric fields affect the parameters of the pseudopotential with straightforward techniques, such as quantum defect theory~\cite{Chin:2010rmp}.

In contrast, for molecules and lanthanide atoms, the rich internal structure leads to a proliferation of closed channel bound states and associated scattering resonances~\cite{Croft:2014pra,Mayle:2012pra,maier:emergence_2015,Frisch2014,Maier:2015pra}. For example, when diatomic molecules collide, there are an enormous number of  rovibrational excitations of the four-atom tetramer complex that forms. Extending previous results~\cite{Forrey:1998pra,avdeenkov:ultracold_2001,bohn:rotational_2002,
tscherbul:controlling_2006,simoni:ultracold_2006,quemener:vibrational_2008,
tscherbul:magnetic_2009,simoni:feshbach_2009,zuchowski:reactions_2010,
croft:multichannel_2011}, mostly on lighter molecules, Refs.~\cite{Mayle:2012pra,Mayle:2013pra} highlighted  this  collisional complexity in the context of bialkali molecules, elucidated many of its properties, and predicted the density of states for molecule-molecule  closed-channel bound states  to  be as high as 1/nK.   In lanthanide atoms the closed-channel bound state density is observed to be  $\sim1$/(100$\mu$K). In addition to lanthanide atoms and molecules, many-resonance collisional complexity has  been predicted to manifest in excited-state alkaline-earth atom collisions~\cite{green:quantum_2016}. Although in ultracold lanthanide atoms the temperature is much less than the resonance spacing, a multichannel treatment of the interactions is still  necessary to describe the dependence on external fields, as well as the cases where resonances overlap. In molecules, a multichannel model is even more essential, since even at the lowest temperatures -- or even at zero-temperature in an optical lattice or tight trap~\cite{Docaj:2016prl,Wall:2017pra,Wall:2017pra2} -- hundreds of channels can be relevant. The two-particle physics remains  under active investigation~\cite{jachymski:impact_2016,Jachymski:2015pra,jachymski:analytical_2013,
yang:observation_2018,dawid:two_2018,makrides:fractal_2018,augustovic:manifestation_2018,
croft:universality_2017,Croft:2017pra,frye:approach_2016,yang:classical_2017}.  Beyond two-particle physics,  multichannel interactions can have dramatic consequences, for example on many-body phase diagrams~
\cite{ewart:bosonic_2018}. Thus an effective interaction capable of treating the multichannel scattering is essential. 

In this paper we present a collisional model  that is simultaneously flexible enough to account for the collisional complexity of systems like ultracold molecules and lanthanide atoms, yet simple enough that one can semi-analytically calculate scattering properties and incorporate them into many-body theories. The model is a multichannel potential, where each channel is a square well with radius $r_0$ and where each pair of channels is coupled by a constant for interparticle separations $r<r_0$, generalizing the two-channel square well model~\cite{Noyola:1977jcp,Chin:2010rmp,Wasak:2014pra,Kokkelmans:2002pra}. This model is one of the simplest finite-ranged alternatives to the effective zero-range multichannel model~\cite{Docaj:2016prl,Wall:2017pra,Wall:2017pra2}, which is more indirectly connected to the physical states  and requires a cumbersome regularization. It is also an alternative to more accurate scattering calculations with multichannel potential energy curves, which are difficult both to accurately calculate and calibrate to experimental data. 

We also present a method to choose parameters for the multichannel square well interaction, which mimic the features of complex ultracold collisions. The spectral statistics of many complex systems may be understood within the framework of random matrix theory (RMT)~\cite{MehtaML:2004random}, which grew out of the study of nuclear spectra~\cite{Weidenmuller:2009rmp,Mitchell:2010rmp,Brody:1981rmp}. The basic premise of RMT is that the spectral statistics of complex systems are reproduced by an ensemble of random matrices.  The statistics are universal, depending only on the symmetries of the underlying Hamiltonian. Most relevant to the present work is the Gaussian orthogonal ensemble (GOE), which describes systems with time-reversal symmetry and draws Hamiltonian matrices from a distribution with probability $P({\mathbf H})\propto \exp{\left(-\text{Tr} {\mathbf H}^2/2\sigma^2\right)}$. The model presented here is inspired by RMT, drawing channel couplings from a Gaussian distribution, but adds a layer of flexibility by incorporating RMT ideas into a more traditional scattering calculation. This flexibility allows---among other things---for independent modeling of the collision thresholds, spatial structure of resonance wavefunctions, and resonance widths.

The paper is laid out as follows. Section~\ref{sec:multichannel-square-well-model} first defines the $N$-channel square-well model. Then it calculates the two-particle scattering eigenstates. Specifically, it reduces finding these eigenstates to a solving a  set of $2N+1$ linear equations for $2N+1$ variables whose coefficients are simple analytic functions (Bessel functions). Section~\ref{sec:scattering-props} relates the scattering properties of interest -- the closed channel fraction $Z_c$, scattering phase shifts $\delta_l$, cross section $\sigma$, Wigner-Smith time delay $\tau$, and the resonance positions $E_R^{(i)}$ and widths $\Gamma^{(i)}$ -- to the multichannel wavefunctions. Section~\ref{sec:pedagogical-example}  
will give a simple pedagogical example of scattering data for a three-channel model, illustrating the connection between the model parameters and the resulting scattering properties.
Section~\ref{sec:statistics} contains a statistical analysis of the spectral data for an ensemble of systems. Section~\ref{sec:statistical-model} introduces the statistical ensemble of specific parameter choices for the multichannel square-well model, and Sections~\ref{sec:sampspectrum} and~\ref{sec:statmeasures} present the resulting scattering data. While these parameter choices are in the spirit of a toy model, they are nevertheless physically realistic, possessing an overall structure similar to that expected for collisionally complex systems. The analysis shows that the multichannel square-well model together with the statistical ensemble of parameters captures the crossover between integrable and chaotic scattering as a function of closed-channel coupling strength in a natural way.  Section~\ref{PhysContext} describes how to determine the multichannel model parameters from physical properties such as scattering data, essential for  using the model as a pseudopotential. Section~\ref{conclusions}  concludes.

\section{Multichannel square-well model \label{sec:multichannel-square-well-model}}

In this section, we present the multi-channel square-well interaction model, and we solve it for the  two particle eigenstates. We first  reduce the $N$-channel problem with one channel open to a system of $2N+1$ linear equations for $2N+1$ variables. These equations all have analytic coefficients, and they can then readily be solved numerically. 

We consider a multichannel two-body Schr\"odinger equation with a central potential in the relative coordinate $r$ of the form
\begin{equation}
\label{eq:mcse}
   \left[{\bf 1} H_0  + {\bf V}(r)\right]\vec{\psi}(r)=E\vec{\psi}(r).
\end{equation}
where ${\bf 1}$ is the identity matrix, and
\begin{equation}
   H_0 = \frac{\hbar^2}{2\mu}\left(-\pdd{}{r} + \frac{l(l+1)}{r^2} \right)
\end{equation}
is the kinetic energy operator in radial coordinates, with $\mu$ the reduced mass colliding for the two particles and $l$ the relative angular momentum. The  wavefunction is a vector, and its components $\psi_i$ represent the wavefunction in channel $i$. 
 
\begin{figure}
\begin{center}
    \includegraphics[width=0.45\textwidth]{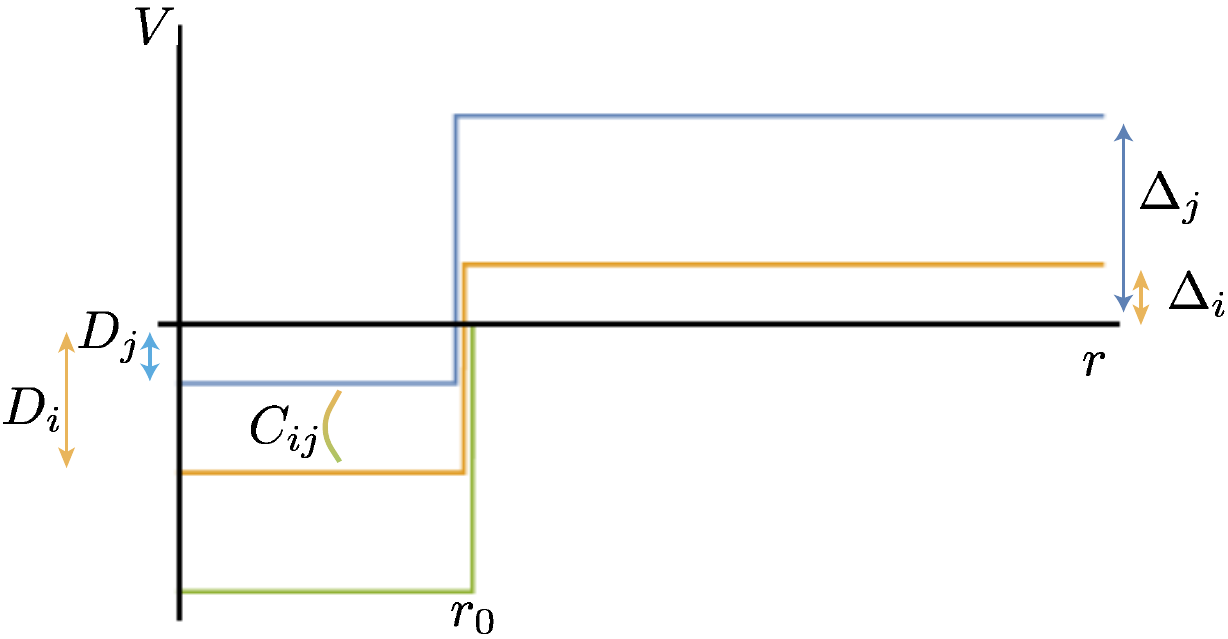}
    \end{center}
    \caption{(color online) Illustration of the multichannel square well model, defined in Eq.~\eqref{eq:mcv}. Small offsets in the square well radius are included only for visual clarity. 
    \label{fig:multichannel-sq-well}  
    }  
\end{figure}

We take the potential matrix to be piece-wise constant, given by
\begin{equation}
\label{eq:mcv}
    V_{ij}(r) = 
    \begin{cases}
    -D_i        & \hspace{0.3in}\left( i = j, r<r_0\right)\\
    C_{ij}      & \hspace{0.3in}\left(i\ne j, r<r_0\right)\\
    \Delta_i \delta_{ij}   &\hspace{0.3in}\left(r>r_0\right),
    \end{cases}         
\end{equation}
illustrated in Fig.~\ref{fig:multichannel-sq-well}.  
We shall restrict our analysis here to the case of only one open channel (for which $E>\Delta_i$) and $N-1$ closed channels (for which $E<\Delta_i$). 
In all calculations that follow, we choose the open channel to be $i=N$ and set this threshold energy to zero, i.e. $\Delta_N = 0$.  

In the region $r>r_0$, the Hamiltonian decouples into  $N$ independent equations, 
\begin{equation}
\label{eq:de1}
    \left(-\pdd{}{r} + \frac{l(l+1)}{r^2} + \frac{2\mu\Delta_i}{\hbar^2} - k^2\right)\psi_i(r)=0
\end{equation}
with $k = \sqrt{2\mu E}/\hbar$, and the solutions in each channel are Riccati-Bessel functions.  
For each channel, two linearly independent solutions are given by
\begin{eqnarray}
\label{eq:rbfunc}
    \{f_i(x),g_i(x)\} = 
    \begin{cases}
      \sqrt{\frac{\pi x}{2}}\left\{J_{l+\frac{1}{2}}(x),N_{l+\frac{1}{2}}(x)\right\} & i \in o\\
      \sqrt{\frac{2 x }{\pi}}\left\{I_{l+\frac{1}{2}}(x),K_{l+\frac{1}{2}}(x)\right\} & i \in c
    \end{cases}        
\end{eqnarray}

For the open channel, $J_\nu(x)\; (N_\nu(x))$ are the Bessel function of the first (second) kind of order $\nu$~\cite{NISThandbook}, $\Delta=0$, and $x\rightarrow k r$.
The normalization of the open channel functions is $\int_0^\infty f_{}(kr)f_{}(k^\prime r)dr={\frac{\pi}{2}}\delta(k-k^\prime)$ for a given $l$.
The closed-channel solutions ($E<\Delta_i$)  involve modified Bessel functions, and 
$x\rightarrow \kappa_i r$ with $\kappa_i = \sqrt{2\mu(\Delta_i-E)}/\hbar$.  In the exterior region $r>r_0$, we retain only the exponentially decaying solution, $g_i(\kappa_ir)$, that vanishes as $r\rightarrow\infty$. Therefore the solution vector $\vec{\psi}^+$ in the exterior region has components
\begin{equation}
\label{eq:exterior-solns}
    \psi_i^{+}(r) = 
    \begin{cases}
    a_i g_i(\kappa_i r) & i\in c \\
    c f_i(k_ir) - s g_i(k_ir) & i \in o
    \end{cases} \hspace{0.27in} \text{for } r>r_0.
\end{equation}
The unknown coefficients $a_i, c, s$ are to be determined.  If we were to choose the open channel amplitude to be normalized so that $c^2+s^2 = 1$, we would find that $s\rightarrow\sin(\delta_l)$ and $c\rightarrow\cos(\delta_l)$ with $\delta_l$ the angular momentum $l$ scattering phase shift.

In the interior region ($r<r_0$), the solution to Eq.~(\ref{eq:mcse}) follows from the fact that the potential matrix ${\bf V}$ is constant, and therefore can be diagonalized by a constant orthogonal transformation, ${\bf U}$, which commutes with $H_0{\mathbf 1}$. We refer to the eigenvalues of ${\bf V}$ as $\epsilon_\alpha$.  They form the diagonal elements of ${\mathbf \Lambda} = {\bf U}^T {\bf V} {\bf U}$.  Inverting the orthogonal transformation, ${\bf V}=  {\bf U} {\bf \Lambda} {\bf U}^T$, we  rewrite Eq.~(\ref{eq:mcse}) in the interior region as
\begin{equation}
\label{eq:dmcse}
     \left({\bf 1}H_0 + {\bf \Lambda}\right){\bf U}^T\vec{\psi}(r) = E{\bf U}^T\vec{\psi}(r).
\end{equation}
Thus Eq.~(\ref{eq:mcse}) is reduced in the interior region to a set of uncoupled equations for the components of $\vec{\phi}={\bf U}^T\vec{\psi}$.  Each solution vector has only one nonzero component: $[\vec{\phi}_\alpha(r)]_{\beta}=\delta_{\alpha \beta}\phi_\alpha(r)$. The component solutions---which are required to vanish at the origin---are again Riccati-Bessel functions. Defining $k^2_\alpha = - \kappa^2_\alpha = 2\mu(E-\epsilon_\alpha)/\hbar^2$, the solutions are
\begin{equation}
\label{eq:int}
    \phi_\alpha(r) = 
    \begin{cases}
    f_\alpha(k_\alpha r) & E > \epsilon_\alpha \\
    f_\alpha(\kappa_\alpha r) & E < \epsilon_\alpha
    \end{cases} \hspace{0.27in} \text{for } r<r_0
\end{equation}
One can easily rotate the solution back to the original channel basis, however the result $\vec{\psi}_\alpha={\mathbf U}\vec{\phi}_\alpha$ does not in-general match smoothly onto the exterior solutions given in Eq.~(\ref{eq:exterior-solns}).  The physical interior solutions must be constructed by taking linear combinations of $\vec{\psi}_\alpha$ with coefficients $b_\alpha$ to be determined by the matching condition. This means the interior wavefunction will be written as $\vec{\psi}^- = \sum_\alpha{b_\alpha \vec{\psi}_\alpha}$. Writing out the $i$ component of $\vec{\psi}^-$ we have:
\begin{equation}
\label{eq:interior-solns}
 \psi_i^{-}(r)=\sum_\alpha{b_\alpha U_{i\alpha}\phi_\alpha}(r)\hspace{0.27in} \text{for } r<r_0.
\end{equation}
It is convenient to define the $i$ component of $\vec{\psi}_\alpha$ as
\begin{equation}
    \Phi_{i\alpha}(r)=U_{i\alpha}\phi_\alpha(r).
\end{equation} 

We now express the requirement that the wavefunction and its derivative be continuous at $r_0$ for every physical channel: $i=\{o,c\}$ as
\begin{eqnarray}
\label{eq:bc}
\sum_\alpha{b_\alpha \Phi_{i \alpha}(r_0)} &&=    \psi_i^{+}(r_0) \\
\sum_\alpha b_\alpha {\pd{\Phi_{i \alpha}(r_0)}{r}} &&= \pd{\psi_i^{+}(r_0)}{r}. \nonumber
\end{eqnarray}

There are now $N$ unknown coefficients $b_\alpha$ that characterize the appropriate linear combination in the interior region, along with $N+1$ unknowns $\{a_i,c,s\}$ with $i=1$, $2$, $\ldots$, $N-1$ where $c$ and $s$ specify the overall normalization and phase shift of the open-channel wavefunction. With only one entrance channel, the solution is unique up to the normalization of the open channel. We first exploit this remaining freedom in the overall normalization by setting any one of the $\{a_i,c,s\}$ to unity, and then normalize the wavefunction according to $\vec{\psi}\rightarrow \vec{\psi}/\sqrt{c^2+s^2}$.  Collecting the $2N+1$ unknowns into a single vector $\vec{x}$ with elements $\{x_j\}=\{b_\alpha,a_i,c,s\}$, we can now set up a linear system of equations of the form ${\bf A}{\vec x}=\vec{y}$. The matrix ${\bf A}$ will be composed of the wavefunctions 
in Eq.~(\ref{eq:bc}) collected in a $(2N+1)\times(2N+1)$ matrix. 
Then $\vec{y}$ and the last row of ${\bf A}$ will be a vector with one non-zero entry corresponding to our arbitrary choice of overall wavefunction amplitude, such as $c=1$.
This leaves a total of $2N$ unknowns that can be determined by the remaining $2N$ equations that express continuity of the wave functions and their derivatives at $r=r_0$. Finally, the problem is expressed as:
\begin{widetext}
\begin{equation}
\label{eq:Axb}
 \left(
\begin{array}{cccccccc}
 \Phi _{11} & \cdots  & \Phi _{1 N} & -g_1 & ... & 0 & 0 & 0 \\
 \frac{\partial \Phi _{11}}{\partial r} & \cdots  & \frac{\partial \Phi _{1 N}}{\partial r} & -\frac{\partial g_1}{\partial r} &  & 0 & 0 & 0 \\
 \vdots  & & \vdots  &  & \ddots &  &  & \vdots \\
 \Phi _{N-1,1} & \cdots  & \Phi _{N-1,N} & 0 &  & -g_{N-1} & 0 & 0 \\
 \frac{\partial \Phi _{N-1,1}}{\partial r} & \cdots  & \frac{\partial \Phi _{N-1,N}}{\partial r} & 0 &  & -\frac{\partial g_{N-1}}{\partial r} & 0 & 0 \\
 \Phi _{\text{N1}} & \cdots  & \Phi _{\text{NN}} & 0 & & 0 & -f_N & g_N \\
 \frac{\partial \Phi _{\text{N1}}}{\partial r} & \cdots  & \frac{\partial \Phi _{\text{NN}}}{\partial r} & 0 &  & 0 & -\frac{\partial f_N}{\partial r} & \frac{\partial g_N}{\partial r} \\
 0 & 0 & 0 & 0 & ... & 0 & 1 & 0 \\
\end{array}
\right)
\left(
\begin{array}{c}
 b_1 \\ \vdots  \\ b_N \\ a_1 \\ \vdots  \\ a_{N-1} \\ c \\ s \\
\end{array}
\right)
=\left(
\begin{array}{c}
 0 \\ 0 \\\vdots  \\ 0 \\ 0 \\ 0 \\ 0 \\ 1 \\
 \end{array}
\right).
\end{equation}
\end{widetext}
This equation can be solved with a standard linear algebra package and its 
solution -- a vector containing $b_i$, $a_i$, $c$, and $s$ -- can be used to construct the 
complete scattering solution.

\section{Scattering properties and observables\label{sec:scattering-props}}

All physical quantities of interest may be extracted from the energy-dependent solution vector $\{b_i,a_i,c,s\}$. Here, we describe the observables calculated in this work, briefly discuss their utility, and outline how they are determined from the solutions to Eq.~\eqref{eq:Axb}.

\subsection{Observables}
One quantity calculated directly from the wavefunctions is the closed-channel population,
\begin{equation}
    Z_C=\sum_{i\in c}\int_0^\infty |\psi_i(r)|^2 dr.
\end{equation}
This quantity is a measure of how much probability density resides in a closed-channel quasi-bound state.  This can be easily calculated with a numerical quadrature grid.
It is sharply peaked at resonance energies and can therefore be used to identify resonance positions.
This quantity can be probed with photoassociation, and has been used in experiments probing the many-body BEC-BCS crossover~\cite{partridge2005}.

Another important scattering observable, from which we can calculate the rest of the other scattering properties of interest here, is the scattering phase shift $\delta_l$. This is defined by
\begin{equation}
\label{eq:tandel}
    \tan(\delta_l) = \frac{s}{c}.
\end{equation}
From this, one can calculate
the scattering cross section (assuming s-wave only with $\delta = \delta_{l=0}$), 
\begin{eqnarray}
\label{eq:sigma}
\sigma=\frac{4 \pi}{k^2} \sin^2(\delta),
\end{eqnarray}
and the Wigner-Smith time delay~\cite{Wigner:1955PR,Smith:1960PR},
\begin{equation}
\label{eq:td}
    \tau = 2\hbar \pd{\delta}{E}.
\end{equation}
Resonance positions are identified by searching for maxima in the time delay.  This procedure identifies the real part $E_R^{(i)}$ of the corresponding pole in the scattering matrix. It does not typically coincide with the maximum value of the cross section, which is more directly related to the magnitude of the pole~\cite{LunaAcosta:2016PhysLettA,Klaiman:2010JPhysB}.
The value of the time delay at the resonance position is in turn related to the width of the resonance~\cite{friedrich:2013scattering}:
\begin{equation}
    \Gamma^{(i)} = \frac{4\hbar}{\tau(E_R^{(i)})},
\end{equation}
The square of the scattering amplitude, $\sin^2{(\delta)}$ in the vicinity of an isolated resonance is well described by the Fano line shape~\cite{fano:1986atomic,rau2004:physicascripta}:
\begin{equation}
    \label{eq:fanocurve}
    \sin^2{(\delta)}=\sin^2{(\delta_{bg})}\frac{\left(E-E_R^{(i)}+q\frac{\Gamma^{(i)}}{2}\right)^2}{\left(E-E_R^{(i)}\right)^2+\left(\frac{\Gamma^{(i)}}{2}\right)^2}.
\end{equation}
Here, $\delta_{bg}$ is the background scattering phase shift due to the open channel only:
\begin{equation}
    \tan{\delta_{bg}}=\frac{\gamma f_o(k r_0) -  k f_o^\prime(k r_0)}{\gamma g_o(k r_0) - k g_o^\prime(k r_0)},
\end{equation}
where $k=\sqrt{2\mu E}/\hbar$ and $\gamma=k_{\text{in}}f_o^\prime(k_{\text{in}}r_0)/f_o(k_{\text{in}}r_0)$ with $k_{\text{in}} = \sqrt{2\mu(E + D_{N})}/\hbar$.  The Fano $q$ parameter is determined by $\delta_{bg}$:
\begin{equation}
\label{eq:fanoq}
    q=-\cot{(\delta_{bg})}
\end{equation}
No ``fitting procedure"  is needed to correctly reproduce the Fano lineshape for each resonance --- all parameters needed to evaluate Eq.~(\ref{eq:fanocurve}) are determined at the resonance energy $E_R^{(i)}$. 

\subsection{Time Delay}
The time delay Eq.~(\ref{eq:td}) is of paramount interest in the determination of resonance positions and widths. One strategy to calculate $\tau(E)$ would be to calculate $\delta(E)$ and apply a numerical derivative with respect to energy. This procedure, however, is limited by the accuracy of the numerical derivative, and moreover makes searching for the maxima of the time delay an inelegant procedure. It is possible, however, to avoid the numerical derivative and calculate the time delay and its energy derivative (necessary to determine $\tau)$, directly at a given energy.  
We proceed by adapting a strategy employed for $R$-matrix methods~\cite{Walker:1989JCP}. Begin with Eq.~(\ref{eq:Axb}), which is of the form 
\begin{equation}
\label{eq:Axb-abs}
    {\bf A}\vec{x}=\vec{y},
\end{equation} and apply an energy derivative to both sides. 
 Since the vector $\vec{y}$ is a constant, we immediately obtain a linear equation for $\der{\vec{x}}{E}$:
\begin{equation}
\label{eq:tdeq}
    {\mathbf A}\der{\vec{x}}{E}=-\der{\mathbf A}{E}\vec{x}
\end{equation}
Let us consider the energy derivative of the matrix ${\bf A}$. Because the matrix ${\bf U}$ is independent of the energy,  $\der{\Phi_{i\alpha}}{E}=U_{i\alpha}\der{\phi_\alpha}{E}$, and hence all energy derivatives are applied to 
Riccati-Bessel functions of the form given in Eqs.~(\ref{eq:rbfunc}). The second derivative $\der{}{E}\left(\der{\phi_\alpha}{r}\right)$ may be efficiently evaluated without calculating any additional functions by invoking the Riccati-Bessel differential equation:
\begin{equation}
    \secder{f_l^{c,o}(z)}{z}=\left(\frac{l(l+1)}{z^2}\pm 1\right)f_l^{c,o}(z),
\end{equation}
 where we take the $+$ sign for the exponential functions ($c$) and the $-$ sign for the oscillatory ($o$) functions.  The vector $\vec{x}$ appearing in Eq.~(\ref{eq:tdeq}) is the result of solving Eq.~(\ref{eq:Axb-abs}).  The solution to Eq.~(\ref{eq:tdeq}), namely $\der{\vec{x}}{E}$, gives $\der{s}{E}=\der{x_{2N+1}}{E}$ and $\der{c}{E}=\der{x_{2N}}{E}$. 
 These, in turn are related to the time delay through Eq.~(\ref{eq:tandel}) and Eq.~(\ref{eq:td}). 
 \begin{equation}
 \label{eq:tdformula}
     \tau(E) = \frac{2\hbar \cos^2(\delta)}{c^2}\left(c \der{s}{E} - s \der{c}{E}\right) 
 \end{equation}
If, as in Eq.~(\ref{eq:Axb}), one chooses $c(E) = 1$, then the Eq.~(\ref{eq:tdformula}) simplifies to $\tau(E)= 2\hbar \cos^2(\delta) \der{s}{E}$. Because $\tau(E)$ exhibits a peak at resonance, the search for resonances is equivalent to a search for the zeroes of $\der{\tau}{E}$. To calculate $\der{\tau}{E}$ we apply another energy derivative to Eq.~(\ref{eq:tdeq}) and solve the resulting linear equation:
\begin{equation}
    {\bf A}\secder{\vec{x}}{E} = -2 \der{{\bf A}}{E}\der{\vec{x}}{E} - \secder{\bf A}{E}\vec{x}
\end{equation}
It is again possible to efficiently evaluate the elements of $\secder{\bf A}{E}$ without calculating any additional Riccati functions beyond those needed for ${\bf A}$. If we again choose $c(E)=1$, the derivative of the time delay is 
\begin{equation}
\der{\tau}{E} = 2\hbar \cos^2(\delta)\secder{s}{E} - \frac{s\tau^2}{\hbar}
\end{equation}
In this way, each additional energy derivative of the solution vector may calculated at the cost of solving only one additional linear matrix equation at the same energy.

\subsection{Bound State Sector}
When the coupling between the open and closed channels is weak, the open channel serves as an ``analyzer" for the spectrum of the closed-channel sector. A clear understanding of the elastic scattering spectrum emerges if one first calculates the position of the closed-channel bound states. 
Let us therefore restrict our attention for the moment to the sector of closed channels only, considering the $N_c = N-1$ closed channels to be isolated from the open channel. 
One must now return to the potential matrix and diagonalize only the closed-channel sector of ${\bf V}$.  Let ${\bf \Lambda}_c={\bf U}_c^T{\bf V}_c{\bf U}_c$, where ${\bf V}_c$ is comprised of the first $N_c$ rows and columns of ${\bf V}$. We may then construct a matrix of the functions $\Phi_{i\alpha}^c = U_{i \alpha}^c\phi_\alpha^c$. 
Then, we match the log-derivatives of the interior and exterior solutions by demanding:
\begin{equation}
    \frac{\sum_\alpha b_\alpha^c \pd{\Phi_{i \alpha}^c(r_0)}{r}}{\sum_\alpha{b_\alpha^c \Phi_{i \alpha}^c(r_0)}} =   \frac{1}{g_i(\kappa_i r_0)}\pd{g_i(\kappa_i r_0)}{r} 
    \text{ for } i \in c.
\end{equation}
We can write this as a matrix equation if we let ${\bf G}$ be an  $N_c\times N_c$ matrix whose diagonal elements are the log-derivatives of the exterior functions
\begin{equation}
G_{ij}=\delta_{ij}\left[\frac{1}{g_i(\kappa_i r)}\der{g_i(\kappa_i r)}{r}\right]_{r_0}.
\end{equation}
Now let ${\bf \Phi^c}$ each be $N_c\times N_c$ matrices evaluated at $r_0$ with elements $\Phi_{i\alpha}(r_0)$.
Matching the log-derivatives of the interior solution to the exterior solution leads to a matrix equation
\begin{equation}
\label{eq:bsdet1}
  \left( \pd{{\bf \Phi}}{r}-{\bf G}{\bf \Phi}\right)\vec{z}=0
\end{equation}
where $\vec{z}$ is a vector of containing the $N_c$ coefficients $b_\alpha^c$.  
Equation~(\ref{eq:bsdet1}) is satisfied when
\begin{equation}
\label{eq:bsdet2}
    C(E)=\det{\left| \pd{{\bf \Phi}}{r}-{\bf G}{\bf \Phi}\right|}=0
\end{equation}
The positions of the closed-channel bound states coincide with zeroes of $C(E)$.  If the couplings between the open channel and closed channels are weak, then the zeroes of $C(E)$ will coincide with the resonance positions. 

\section{Simple Example}
\label{sec:pedagogical-example}

\begin{figure}
\begin{center}
    \includegraphics[width=0.45\textwidth]{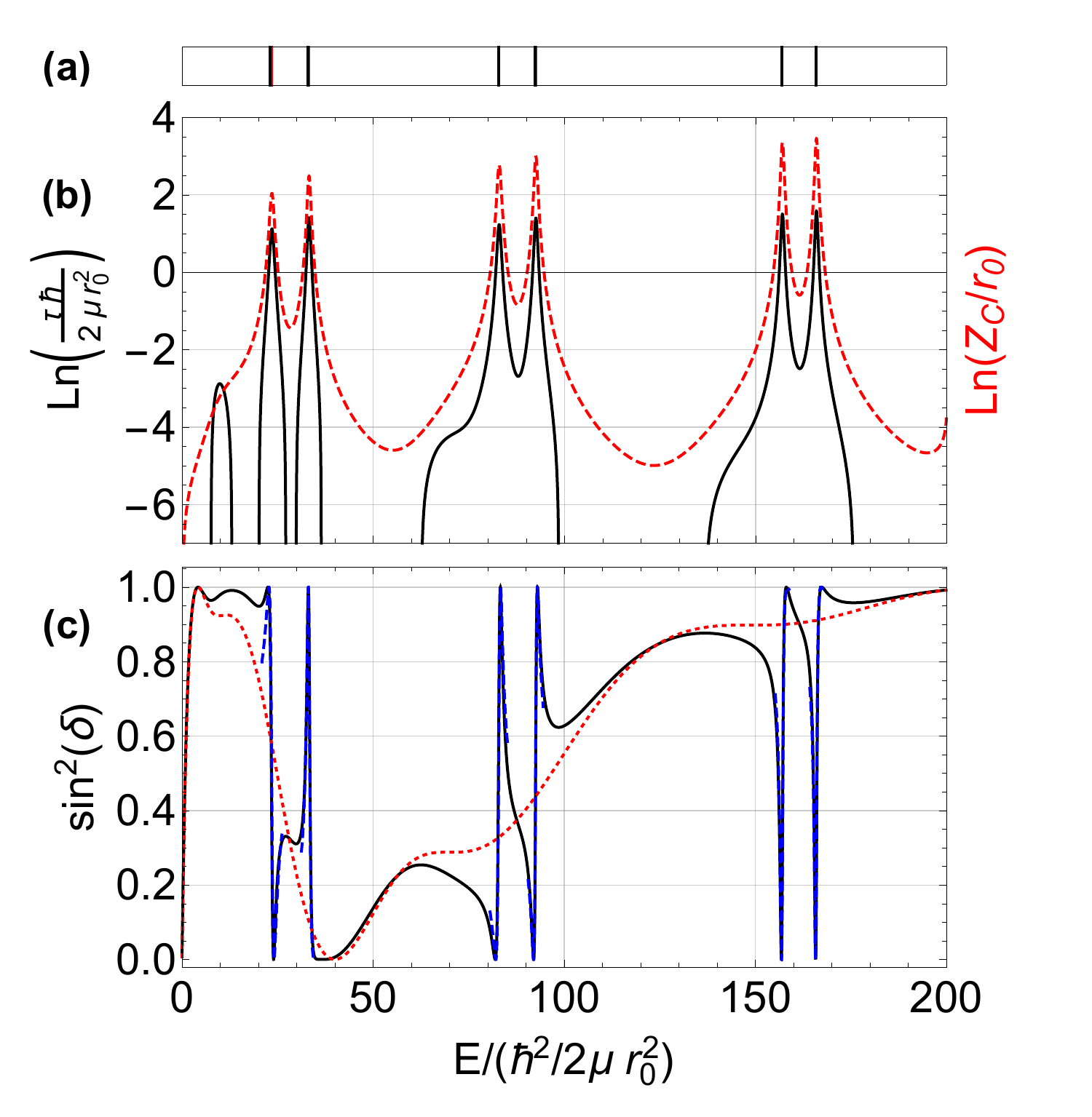}
    \end{center}
    \caption{(color online) Results shown here are for an s-wave, $N=3$ channel scattering systems.
    Panel (a) has black vertical lines indicating the bound states of only the closed channel system, solutions to Eq.~(\ref{eq:bsdet2}).
    The red vertical lines in (a), which lie nearly on top of the black lines,  are the locations of resonances in this system identified as maxima of the time delay. 
    Panel (b) shows the natural-log of both the time delay (black curve) and the closed channel population (red dashed curve).  Sharp peaks occur at the resonance positions. 
    Panel (c) shows $\sin^2(\delta)$ (black). The dashed red curve in panels (c)  corresponds to the background scattering for which the open channel is decoupled from the closed channel sector.
    The blue dashed curves are Fano line shapes plotted using Eq.~(\ref{eq:fanocurve}). 
    \label{fig:scatex}  
    }  
\end{figure}

\begin{figure}
\begin{center}
    \includegraphics[width=0.5\textwidth]{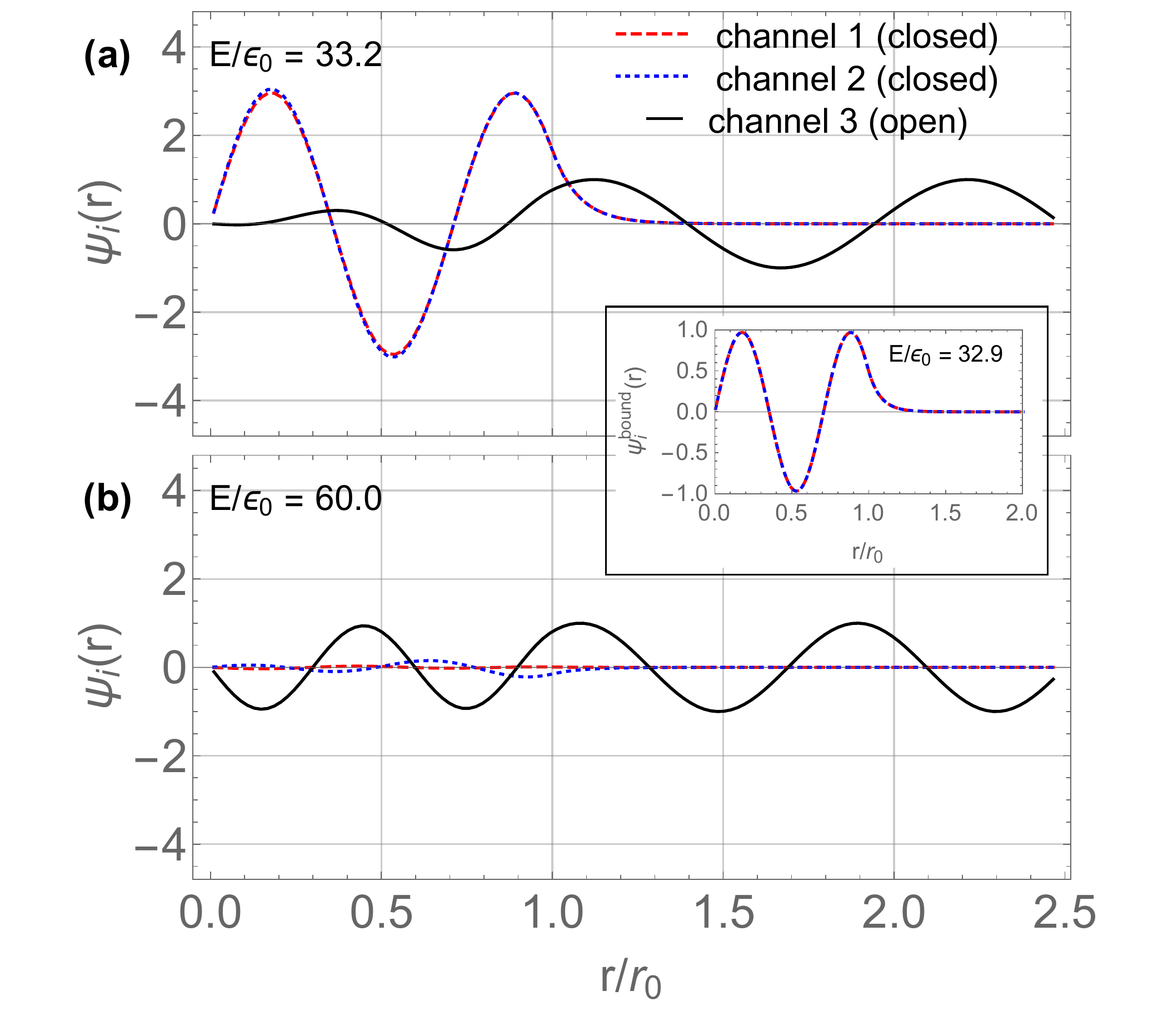}
    \end{center}
    \caption{(color online) The wavefunction is shown for the simple s-wave, $N=3$ channel scattering systems when the systems is both (a) resonant and (b) non-resonant.
    The closed channels are shown are dashed red and dotted blue curves, and the open channel is shown as solid black line. 
    For the scattering wavefunctions, we use the normalization of $s^2+c^2=1$.
    (a) Shows the scattering wavefunction with an energy of 33.2 $\epsilon_0$. 
    (b) Shows the scattering wavefunction with an energy of 60.0  $\epsilon_0$. 
    (Inset) Bound state for the closed sector is shown with energy of 32.9 $\epsilon_0$. 
    \label{fig:psi3}  
    }  
\end{figure}

In this section we present a simple pedagogical example with only three channels which is straightforward to interpret and reproduce.
Before we specify the parameters of the potential matrix, we recall that an isolated (s-wave) square-well potential of depth $V_0$ with respect to its threshold supports bound states with binding energy $B$ that
satisfy the transcendental equation, 
\begin{equation}
\label{eq:t0} 
    \sqrt{\frac{B}{\epsilon_0}} + \sqrt{\frac{V_0-B}{\epsilon_0}}\cot{\left(\sqrt{\frac{V_0-B}{\epsilon_0}}\right)}=0.
\end{equation}
We write energy quantities in terms of the natural energy unit 
\begin{equation}
\label{eq:eps0}
\epsilon_0=\frac{\hbar^2}{2\mu r_0^2}.
\end{equation}   
This notation is used throughout. In the absence of any off-diagonal couplings $C_{ij}\rightarrow 0$, the positive energy ($B<\Delta$) solutions to Eq.~(\ref{eq:t0}) signify bound states embedded into the continuum of the open channel. 

To better illustrate how these bound states become resonances, we consider a tri-diagonal potential matrix with one open channel and two closed channels. The potential is defined by specifying the constants in Eq.~(\ref{eq:mcv}); we choose
 $D_1=D_2=D_3=50\epsilon_0$, $C_{12}=C_{21}=5\epsilon_0$, $C_{23}=C_{32}=5\epsilon_0$, $C_{13}=C_{31}=0$,  $\Delta_1=\Delta_2=200\epsilon_0$ and $\Delta_3=0$.
Each of the two closed channels supports five bound states below its threshold, but only three of these sit above the open-channel threshold. Because both closed channels have the same threshold energy and the same depth, each of the three levels is two-fold degenerate in the limit that all $C_{ij}\rightarrow 0$. The nonzero $C_{ij}$ split these states into six observable resonances with finite width.  The splitting is of order $10\epsilon_0$, as one might expect from inspection of $2\times 2$ closed-channel sector of the potential matrix ${\mathbf V}$.

These six resonances are seen in the scattering observables plotted in Figure~\ref{fig:scatex}. In (a) we compare the resonance positions (red) and the eigenenergies of the bound sector (black). 
In (b) we show both the time delay (black) and closed-channel amplitude ($Z_C$, red). Here we see strong correlations
between the two curves. In (c) we plot $\sin^2(\delta)$ (black) for the full system. Resonances appear clearly upon comparing to the background scattering from only the open channel (red dashed line).
The dashed blue curves in (c) are the Fano resonance profiles with parameters determined from
Eq. (\ref{eq:fanocurve}). Each Fano resonance profile is plotted over the energy range of $E_R^{(i)}-2\Gamma^{(i)}$ to $E_R^{(i)}+2\Gamma^{(i)}$.

In Figure \ref{fig:psi3} we show two wavefunctions for this example system. 
In (a) we show a resonant example for incident energy of 33.2 $\epsilon_0$, and in (b) we shown a non-resonant scattering solution with an energy of 60.0 $\epsilon_0$.  
The closed-channel wavefunctions are shown as dashed red and dotted blue curves, while the open-channel wavefunction is shown as a solid black line. For these wavefunctions, we use the normalization of $s^2+c^2=1$.
The contrast between (a) and (b) is stark. The amplitude of the closed channels are orders of magnitude apart.

In Figure \ref{fig:psi3} the inset shows the 2-channel closed-sector bound state with an energy of about 32.9 $\epsilon_0$. It is this bound state that leads to the resonance shown in (a). This state corresponds to the symmetric linear combination of channel states obtained by diagonalizing the closed-channel sector of ${\mathbf V}$.

\section{Statistical Analysis}
\label{sec:statistics}
Stemming from the well-verified, but yet unproven conjecture of
Bohigas, Giannoni, and Schmit~\cite{Bohigas:1984prl}, the transition
from a regime of  integrability to one of nonintegrability
and chaos is marked by a distinct change in the distribution of energy
level spacings. An integrable system with many degrees of freedom
produces a uniform distribution of energy levels, as if the levels
were the result of a Poisson process giving a Poisson distribution of
energy level spacings.    Because the Hamiltonian for an integrable
system decouples into independent degrees of freedom, the energy
levels may be close together and may even cross as some parameter in
the Hamiltonian is changed.  A chaotic system is typically
characterized by strong coupling between the many degrees of freedom,
leading to a characteristic level repulsion. The nature of the level
repulsion is universal,  depending only on the symmetry. At small level spacings $s$,  the
level spacing probability behaves as $s^{\beta}$ for $\beta = 1, 2$,
or $4$.  We consider only systems with time-reversal symmetry, which 
belong to the $\beta = 1$ universality class.  The corresponding
probability distribution of nearest-neighbor level spacings is to a 
very good approximation described by the Wigner surmise ${\mathcal P}(s)=\frac{\pi s}{2}\exp{\left(-\frac{\pi s^2}{4}\right)}$, also called the Wigner-Dyson (WD) distribution. Here, $s$ is the energy level spacing measured here in units of the
average level spacing: $s = S/\langle S \rangle$.  For experimental data, $\braket{S}$ typically denotes the level spacing averaged over a long spectral run.  For the present calculations, it shall denote both a spectral average and an ensemble average.  The Wigner surmise emerges from the level spacing statistics of an ensemble of $2\times 2$ orthogonal Hamiltonian matrices with elements drawn from a Gaussian distribution. The Gaussian Orthogonal Ensemble (GOE) constitutes a generalization of this idea to an ensemble of $N\times N$ matrices~\cite{MehtaML:2004random}.

Our goal in this section is to introduce a model with physically realistic choices of parameters, and demonstrate not only that it captures the crossover from the Poisson to the Wigner-Dyson regime, but that it does so in a smooth and
physically transparent way. We shall study an ensemble of systems for which the matrix elements of the potential ${\mathbf V}$ between different channels are drawn from
probability distributions inspired by random matrix theory~\cite{Mitchell:2010rmp,Weidenmuller:2009rmp}, but for which each channel has a randomly chosen threshold drawn from a uniform distribution, similar in spirit to the matrix ensembles studied by Wigner~\cite{wigner:characteristic_1955,wigner:characteristics_1957}.  We shall
focus on two physical observables that emerge from this model: the nearest-neighbor
level spacing distribution and the resonance number variance in an energy window.

A convenient probability distribution to describe the nearest-neighbor 
level spacing distribution across the integrable-chaotic crossover 
is the Brody distribution~\cite{Brody:1973LettNuovoCimento, Brody:1981rmp},
\begin{equation}
\label{eq:pbrody}
   \mathcal{P}(w,s)= \left(1+w\right)A(w)s^w \exp{\left(-\left[A(w)s\right]^{1+w}\right)},
\end{equation}
where $A(w)=\Gamma\left(\frac{2+w}{1+w}\right)$ and $\Gamma(x)$ is the Gamma function. 
This smoothly interpolates from the Poisson distribution to the WD distribution
as a single parameter, $w$,  varies from zero to one. The GOE in the limit $N\rightarrow \infty$ gives a Brody parameter $w=0.953$~\cite{Brody:1981rmp}, which is close to, but not exactly equal to one. 

Other statistical measures are often used to characterize fluctuations in the spectral density. In particular, the
number variance, $\Sigma^2$ contains all information regarding two-point fluctuations in the spectrum, and is used to characterize the spectral rigidity.  It is defined as:
\begin{equation}
\label{eq:sigmadef}
    \Sigma^2(\epsilon) = \langle N(E,\epsilon)^2 \rangle - \langle N(E,\epsilon)\rangle^2
\end{equation}
where $\epsilon=\Delta E/\langle S\rangle$ 
and $N(E,\epsilon)$ is the
number of resonances in range $[E,E + \Delta E]$. The bracket $\braket{\cdot}$ in Eq.~(\ref{eq:sigmadef}) may indicate an average over non-overlapping energy windows with starting value $E$, an ensemble average, or---as in our case---both.
The expected number of levels in energy window $\epsilon$ is $\langle N(E,\epsilon)
\rangle \pm \sqrt{\Sigma^2(\epsilon)}$. 
The spectral rigidity itself, typically denoted $\Delta_3$, and its ensemble average $\langle \Delta_3 \rangle$, can be expressed in terms of the number variance (see, for example Ref.~\cite{Brody:1981rmp}). 

\subsection{Model for Potential Matrix Parameters \label{sec:statistical-model}}
Let us first describe our RMT-inspired choice for the potential matrix
${\mathbf V}$, and then construct an ensemble of systems of this form whose
statistical properties we characterize in detail. The model is
required to support a large density of states within a 
prescribed energy window near the collision threshold.  Energy levels
away from threshold are not accessible in ultracold collisions, so our
goal is to construct an optimized model that places levels only where
needed. 
\begin{figure*}[!t]
\begin{center}
    \includegraphics[width=0.45\textwidth]{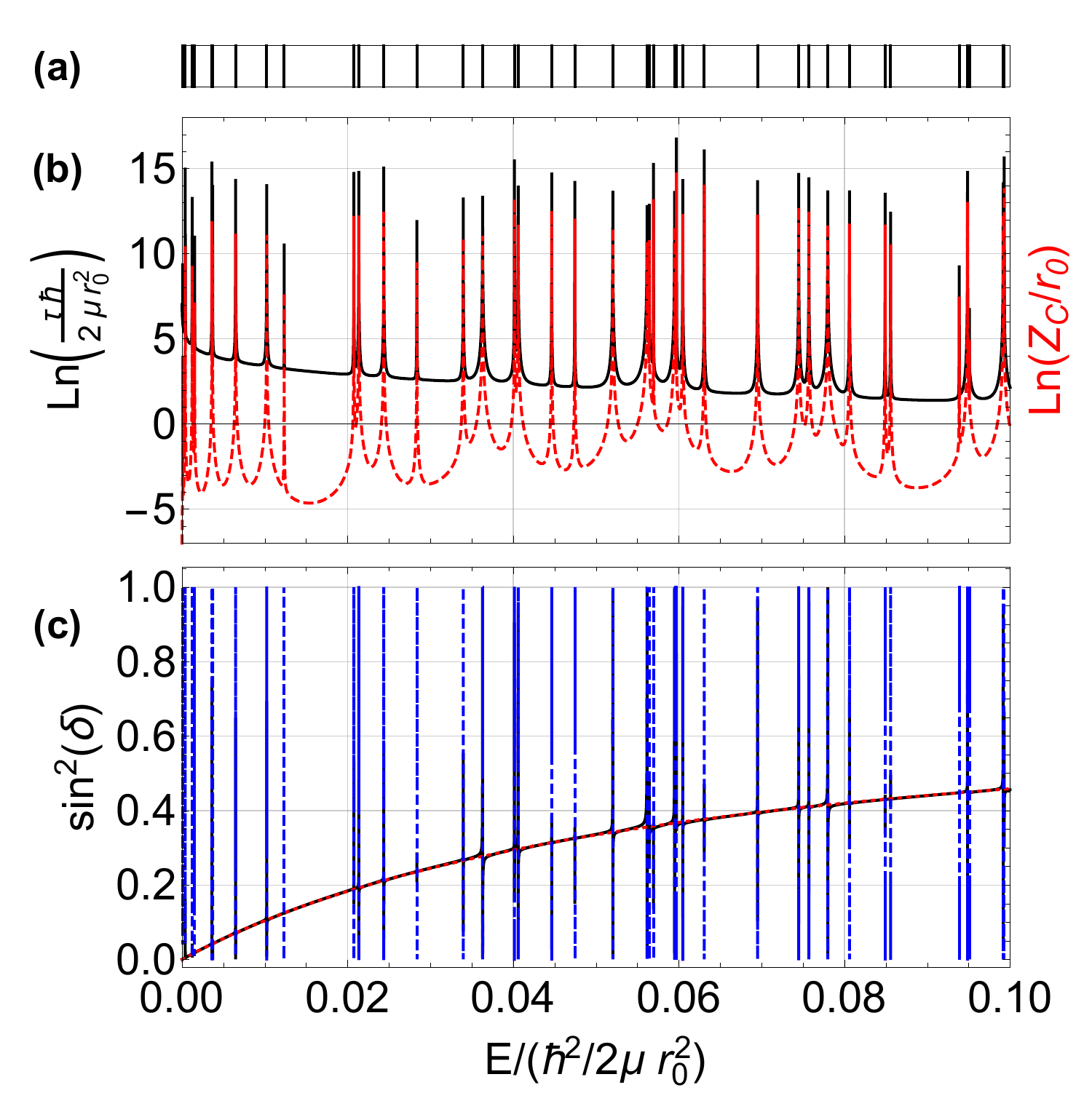}
     \includegraphics[width=0.45\textwidth]{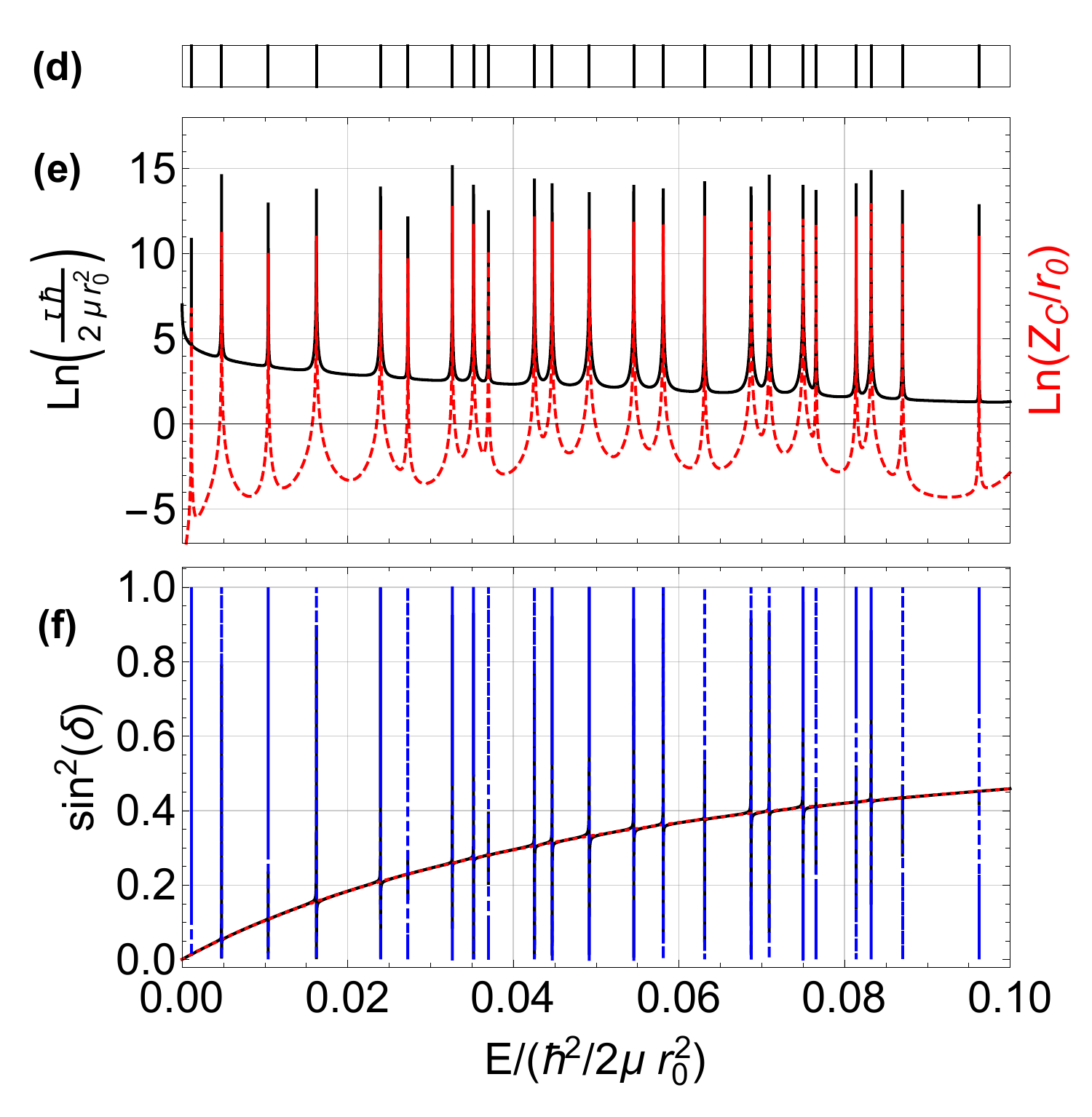}
    \end{center}
    \caption{(color online) Results shown here are for two s-wave $N=41$ channel
      scattering systems. 
    Panels (a-c) show scattering data for a system belonging to an ensemble with a Brody parameter $w = 0.05 \pm 0.01$, indicating an uncorrelated distribution of resonances.  
    Panels (d-f) show data for the same system after the closed-channel coupling has been increased to produce an ensemble Brody
    parameter of $w = 0.96 \pm 0.03$, indicating strong level
    repulsion and chaos.   
    Panels (a) and (d) have vertical lines indicating positions of closed-channel bound states found as solutions to Eq.~(\ref{eq:bsdet2}). 
    Panels (b) and (e) show the natural-log of the time delay (black
    curves) where sharp peaks occur at the resonance positions.  Also shown
    is the closed-channel fraction (red dashed curve).  
    Panels (c) and (f) show $\sin^2(\delta)$ (black). The dashed red
    curve in panels (c) and (f) corresponds to the background
    scattering for which the open channel is decoupled from the closed
    channel sector. 
    \label{fig:scatdata}  
    }  
\end{figure*}
We engineer the desired bound state structure by working with the simplest case of each closed channel supporting exactly one bound state within a fixed energy window near the collision threshold.
A closed channel with threshold $\Delta$ will support a state at the
open-channel threshold (i.e. zero energy) if the binding of the state 
is equal to the threshold energy of that channel: $B\rightarrow
\Delta$. Let $V_0^{(n)}$ be the solution to Eq.~(\ref{eq:t0}) such
that a state with $n-1$ nodes and binding energy $B_n$ sits at the
zero-energy threshold. One can then use $N_c$ closed channels to model
$N_c$ resonances by choosing parameters in Eq.~(\ref{eq:mcv}) such
that 
\begin{align}
\label{eq:model1}
    D_i & = B_n - V_0^{(n)} + E_0^{(i)} \\ \nonumber
   \Delta_i & = B_n + E_0^{(i)} \\ \nonumber
    C_{ij} &= 
    \begin{cases}
    g_{oc} u_{ij} & (i=o,j\in c,\; \text{or}\; i\in c, j=o)\\ \nonumber
    g_{cc} u_{ij} & (i,j \in c) 
    \end{cases}
\end{align}
where $g_{oc}$ represents the scale of the coupling between the open
channel and each closed channel, and $g_{cc}$ represents the scale of
the couplings among closed channels. 

To use this model, we need to choose the parameters appearing in Eq.~(\ref{eq:model1}), for which we use the following physically motivated ensemble. The basic scheme is to imagine each closed channel as belonging to a set of $N_c$ identical clones, each offset so that its bound state sits at position $E_0^{(i)}$. One samples the $E_0^{(i)}$ from a uniform distribution with the desired resonance density, while
the random variables $u_{ij}$ are drawn from a Gaussian distribution
$\mathcal{P}(u)=\exp(-u^2)/\sqrt{\pi}$. The potential matrix is required to be symmetric, so we first fill the upper-right triangle ($j>i$) of the matrix and then set $C_{ij}=C_{ji}$ for all $j<i$. This model captures the key idea that the diagonal and off-diagonal matrix elements have distinct origins in colliding ultracold matter.

In the limit $g_{cc},g_{oc} \ll \langle S \rangle$, the solutions to
Eq.~(\ref{eq:bsdet2}) coincide with the energies $E_0^{(i)}$. As one
increases $g_{cc}$, the resonance positions move away from the initial
values $E_0^{(i)}$, but remain close to the solutions to
Eq.~(\ref{eq:bsdet2}) provided that $g_{oc} \ll \langle S \rangle$. 
It is thus possible to construct an arbitrary density of resonances
within a finite energy window $W_c$ by distributing the $N_c$
resonances $E_0^{(i)}$ as desired within the window $W_c$. We assume
here that the window $W_C$ is smaller than the spacing between the eigenenergies of an individual channel potential, i.e. $W_C<\delta E$,
where $\delta E$ is the spacing between neighboring solutions to
Eq.~(\ref{eq:t0}).  This assumption is not fundamental to the utility
of the model presented here, but it makes the engineering of resonance
states conceptually straightforward. 

All of the data presented in the following subsections are extracted from an ensemble of 100 systems, each with $N=41$ channels, $N_c=40$ of which are closed.  For each system, we place $N_c=40$ resonances within a narrow energy window near the zero-energy collision threshold. We will initialize the ensemble with weak couplings, and consequently the solutions to Eq.~(\ref{eq:bsdet2}) will initially coincide with $E_0^{(i)}$, and the resulting Brody parameter for the ensemble will begin very close to zero.  To study the transition to chaos, we monitor the \emph{coupled} resonance positions $E_R^{(i)}$ as we linearly increase the scale of the closed-channel coupling $g_{cc}$ while keeping $g_{oc}$ fixed.

Clearly, one need not choose the parameters according to Eq.~(\ref{eq:model1}) to model an integrable, chaotic, or intermediate system. One could obtain an ensemble of systems for any desired Brody parameter by simply drawing the initial resonance spacings $E_0^{(i+1)}-E_0^{(i)}$ from the corresponding Brody distribution, and setting $g_{cc}\rightarrow 0$ (with $g_{oc}\ll \braket{S}$). Our goal here is to demonstrate that even when the $E_0^{(i)}$ are uniformly distributed (so the level-spacings are Poisson-distributed), the coupled resonance positions $E_R^{(i)}$ smoothly transition to the Wigner-Dyson regime.

\subsection{Sample Spectrum}
\label{sec:sampspectrum}
 We will show example scattering properties using the statistical ensembles  introduced in the prior section. We choose  the $E_0^{(i)}$ from a uniform distribution in energy range $[0,0.1 \epsilon_0]$. The minimum wavelength within this window of collision energies is much longer than the range of the potential: $\lambda_{\text{min}} \approx 20 r_0$.  Therefore, open-channel collision physics is expected to be insensitive to the short-ranged structure of the potential.  The minimum value for the first closed channel threshold is set to $\Delta_1=10\epsilon_0$.   

We show results for various $g_{cc}$ and $g_{oc}$, summarized in Table~\ref{table:couplings}.
Because couplings are defined in terms of model units $\epsilon_0$, but statistical data is more naturally scaled by $\braket{S}$, both scalings are shown in the table. We choose  $g_{oc}$ to be small compared to both $\epsilon_0$ and $\braket{S}$, and study the dependence on $g_{cc}$ while $g_{oc}$ is held fixed.

\begin{table}[!t]
\begin{center}
\resizebox{\columnwidth}{!}
{
\begin{tabular}{|c|c|c||c|c|c|}
\hline
& \multicolumn{2}{|c||}{Model Units} &   \multicolumn{3}{|c|}{Ensemble Units} \\ 
Fig.~\ref{fig:distributions} & $g_{cc}/\epsilon_0$ & $g_{oc}/\epsilon_0$ & $\braket{S}$ & $g_{cc}/\braket{S}$ & $g_{oc}/\braket{S}$ \\
\hline
(a) & $10^{-5}$     &  $10^{-3}$   & $2.445\times 10^{-3}$ & $4.091\times 10^{-3}$ & $0.4091$  \\
(b) & $1.641\times 10^{-3}$  &  $10^{-3}$   & $2.557\times 10^{-3}$ & $0.6418$              & $0.3911$\\
(c) &$10^{-2}$             &  $10^{-3}$   & $4.029\times 10^{-3}$ & $2.482$               & $0.2482$ \\
\hline
\end{tabular}
}
\end{center}

\caption{\label{table:couplings} Model parameters for the three (out of 50) couplings $g_{cc}$ shown in Fig.~\ref{fig:distributions}. The first two columns show the couplings in model units of $\epsilon_0$, while the final two columns are scaled by $\braket{S}$.
}
\end{table}

\begin{figure}[!t]
    \begin{centering}
    \includegraphics[width=0.45\textwidth]{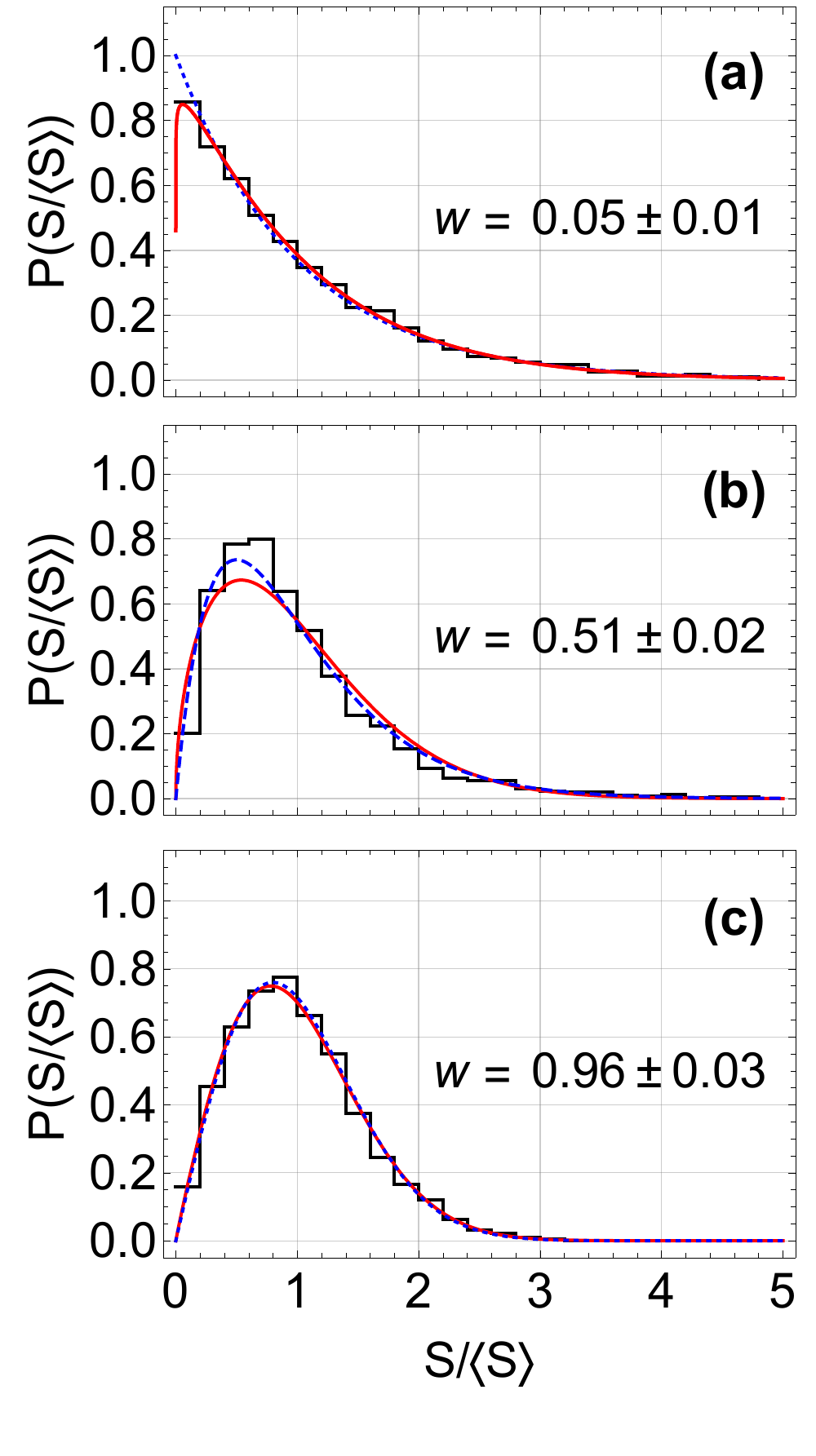}
    \end{centering}
    \caption{ (color online) Three examples of probability distributions of energy
      level spacings calculated from an ensemble of 100 systems each
      of which exhibit spectra like those shown in
      Fig.~\ref{fig:scatdata}.  The coupling values shown for each panel are summarized in Table~\ref{table:couplings}. For each panel, the solid red curve shows the fitted Brody distribution. Panel (a) shows the distribution in
      the regime of Poisson statistics (weak $g_{cc}$).  The dashed blue curve indicates the Poisson distribution. Panel (b)
      shows an intermediate case. The dashed blue curve is a plot of the semi-Poisson distribution.  Panel (c) shows the chaotic limit. The dashed blue curve shows the Wigner-Dyson distribution.
    }\label{fig:distributions}
\end{figure}

Elastic scattering spectra for one particular member of the
statistical ensemble described above are shown in
Fig.~\ref{fig:scatdata}.
Panels (a-c) show data in the regime of Poisson statistics, where the ensemble gives a Brody parameter of $w=0.05\pm 0.01$.  For this case, the couplings are specified by the first row of Table~\ref{table:couplings}.
Panels (d-f) show data in the regime of Wigner-Dyson (or GOE) statistics with
$w=0.96\pm 0.03$. For this case, the couplings are specified by the last row in Table~\ref{table:couplings}.
The finite Brody parameter in the Poisson regime is likely due to the
residual value of $g_{oc}$.  It is possible to reduce the Brody
parameter further by reducing $g_{oc}$.  But the resulting spectra exhibit narrower resonance features, and resolving these  requires calculating scattering data on a finer energy grid. Because we work in the limit of weak $g_{oc}$, the resonance
positions---defined as the maxima of the time delay---are very close
to the positions of the closed-channel bound states given by solutions to Eq.~(\ref{eq:bsdet2}).

The red dashed curve in panels (c) and (f) of Fig.~\ref{fig:scatdata} show the square of the background scattering amplitude, $\sin^2{(\delta_{\text{bg}})}$, due to the open channel only. The background phase shift $\delta_{bg}$ is responsible for the Fano asymmetry parameter $q$ via Eq.~(\ref{eq:fanoq}). Both $\delta_{bg}$ and $q$ can be tuned by varying the open-channel well depth $D_N$. Open-channel bound states near the threshold collision energy have a dramatic effect on the threshold value of $\delta_{bg}$. From Eq.~(\ref{eq:t0}), when $D_N\rightarrow \frac{\hbar^2}{2\mu r_0^2}\left[\frac{(2n-1)\pi}{2}\right]^2$ (for $n=1,2,...$), threshold collisions give $\delta_{bg}\rightarrow \pi/2$ and $q\rightarrow 0$. By varying the depth $D_N$ so that an open-channel bound state crosses the collision threshold, it is possible to model effects similar to the broad resonance features observed in atom loss measurements of dysprosium~\cite{Maier:2015pra}. 

Figure~\ref{fig:scatdata}(b) and (e)   show the time delay (solid black) and the closed channel population (dashed red) as a function of collision energy. Each of these quantities is maximized at resonance, and may be used to identify resonance positions. We locate each maximum in the time delay, and extract the resonance width through Eq.~(\ref{eq:td}).  Note that both $\tau(E)$ and $Z_C(E)$ have maxima that closely coincide with solutions to Eq.~(\ref{eq:bsdet2}) which are marked as vertical lines in panels (a) and (d).

The black curve in panels (c) and (d) of Fig.~\ref{fig:scatdata} shows the square of the total (s-wave) scattering amplitude $\sin^2{(\delta)}$.  It closely follows the background except at resonance positions.  The dashed blue curves are plots of the Fano resonance profile Eq.~(\ref{eq:fanocurve}) from $E_R^{(i)}-2\Gamma^{(i)}$ to $E_R^{(i)}+2\Gamma^{(i)}$.

\subsection{Statistical Measures}
\label{sec:statmeasures}
\begin{figure}
\begin{center}
\leavevmode
    \includegraphics[width=0.44\textwidth]{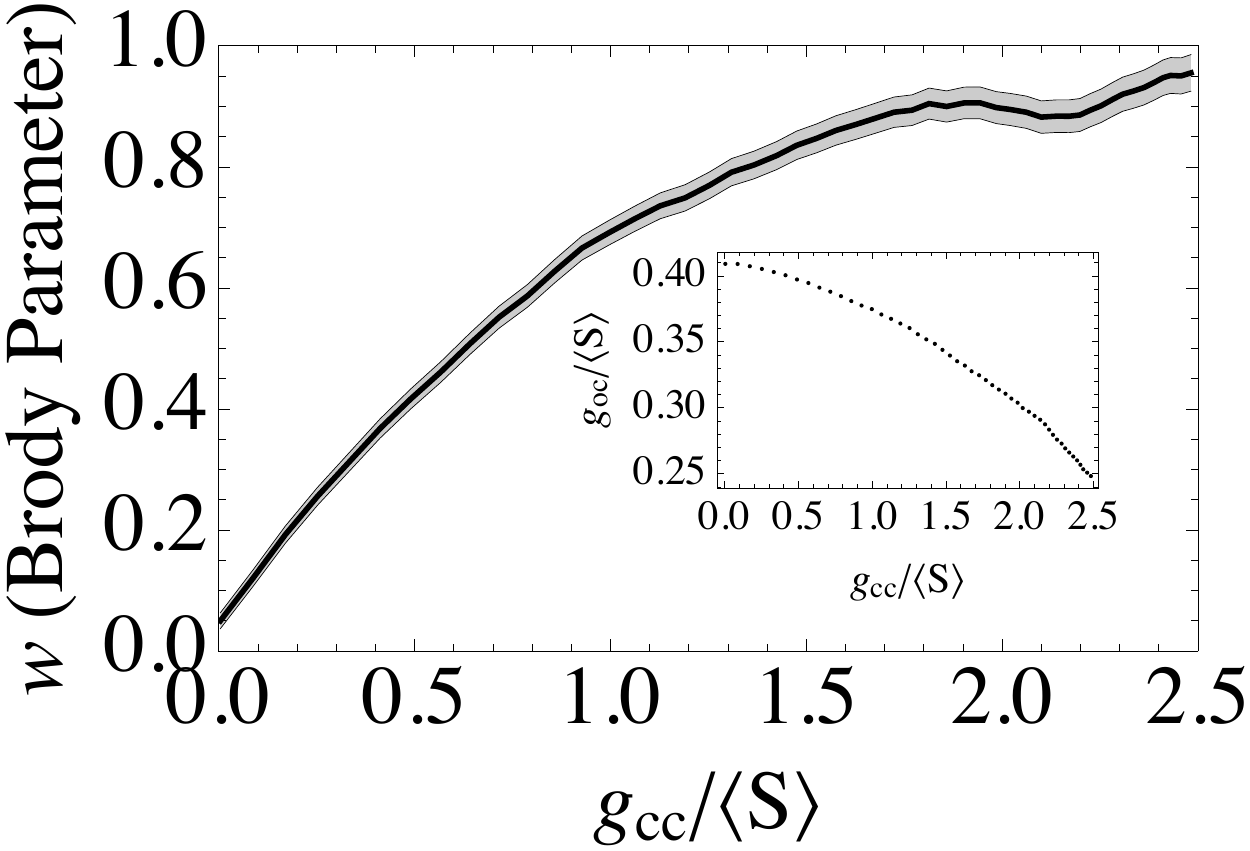}
    \setlength{\unitlength}{\textwidth}
    \end{center}
    \caption{(color online) The Brody parameter $w$ (black) is shown as a function of the closed channel
      coupling strength, which is increased from $g_{cc} = 10^{-5}\epsilon_0$ 
      to $g_{cc}= 10^{-2} \epsilon_0$.  It is rescaled here by the ensemble 
      average energy level spacing $g_{cc}/\langle S \rangle$.  The open-closed
      coupling is held fixed in model units at $g_{oc}=10^{-3}\epsilon_0$.  
      However, due to the repulsion of
      levels, it decreases when scaled by $\langle S \rangle$, as
      shown in the inset.  The shaded region indicates the error calculated using Eq.~(\ref{eq:fiterror}).
    \label{fig:pbrody}
    }
\end{figure}

For each member of the ensemble and for each value of the coupling scale $g_{cc}$, we tabulate $E_R^{(i)}$ in the range $E\in [0,0.1\epsilon_0]$.  From these tabulated values, we calculate ensemble averages reported in this section.
For each coupling, we first compute the average nearest-neighbor level spacing for each system, then average those together to obtain the ensemble averaged $\langle S \rangle$. In Table~\ref{table:couplings} we list the ensemble averaged $\braket{S}$ for a few coupling values.
The level repulsion that occurs as we ramp up the couplings results in an increase in $\langle S \rangle$.  Even when  $g_{oc}/\epsilon_0$ is held is fixed, $g_{oc}/\braket{S}$ decreases with increasing $g_{cc}$, as also shown in the inset of Fig.~\ref{fig:pbrody}.

In order to calculate the Brody parameter that best describes the level spacing distribution for each coupling, we maximize the following log-likelihood function with respect to $w$:
\begin{equation}
\label{eq:loglikelihood}
    \mathcal{M}=\sum_i{\ln{P(w,s_i)}}.
\end{equation}
Here $P(w,s_i)$ is the single-event probability of level spacing $s_i$.  It is given by Eq.~(\ref{eq:pbrody}) up to a multiplicative constant, so $\ln{P(w,s_i)}=\ln{\mathcal{P}(w,s_i)}$ 
up to an irrelevant additive constant. The sum is over all level spacings in the ensemble.  The error in the estimate for $w$ is~\cite{bevington:data_2003}:
\begin{equation}
\label{eq:fiterror}
    \sigma_w=\sqrt{\left[ -\pdd{\mathcal{M}}{w} \right]_{w=w_0}^{-1}}
\end{equation}
where $w_0$ is the value of $w$ that maximizes Eq.~(\ref{eq:loglikelihood}).
In Fig.~\ref{fig:distributions}, we show probability distributions of the nearest-neighbor level spacing for three values of the Brody parameter.  The red curves show the fitted Brody distribution with the corresponding Brody
parameter labeled.  The dashed blue curve in panel (a) shows the Poisson distribution,
\begin{equation}
\label{eq:poissondist}
    {\mathcal P}(s)=\exp{(-s)},
\end{equation}
while the dashed blue curve in panel (b) shows the  ``semi-Poisson" distribution~\cite{GarciaGarcia:2005PRE},  
\begin{equation}
\label{eq:semipoisson}
    {\mathcal P}(s) = 4s \exp{(-2s)}.
\end{equation}
The three panels in Fig.~\ref{fig:distributions} correspond to three points on the crossover curve shown in Fig.~\ref{fig:pbrody}. 

\begin{figure}
\begin{center}
\leavevmode
    \includegraphics[width=0.44\textwidth]{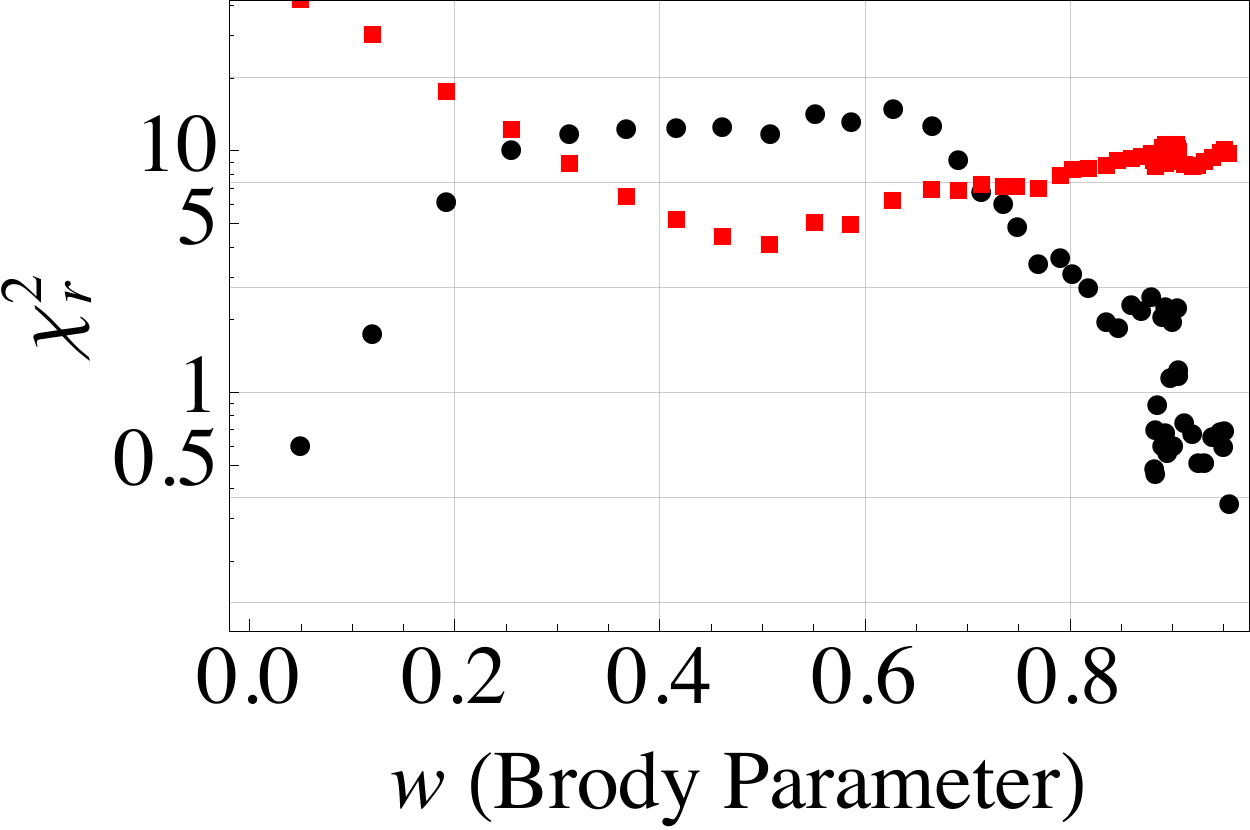}
    \setlength{\unitlength}{\textwidth}
    \end{center}
    \caption{(color online) The reduced chi-squared, $\chi_r^2$, value for the Brody and semi-Poisson distributions are plotted here. We use histograms with 25 bins of size $0.2 \langle S \rangle$, like those shown in Fig.~\ref{fig:distributions}.
    Black circles show $\chi_r^2$ for level-spacing fits to the Brody distribution.  The red squares shows the $\chi_r^2$ value for level-spacing fits to the Semi-Poisson distribution Eq.~(\ref{eq:semipoisson}).  
    \label{fig:chisqr}
    }
\end{figure}

Let us briefly discuss some of the features observed in the crossover
curve shown in Fig.~\ref{fig:pbrody}. We find a rapid rise in $w$ from
zero to about $w=0.7$ as the coupling increases from $g_{cc}\approx 0$
to $g_{cc}\approx \langle S\rangle$, then a more gradual increase with
slow variation towards $w=0.96$ at $g_{cc}\approx 2.5 \langle
D\rangle$. It is unclear at this time to what degree the curve
calculated here for the square-well model Eq.~(\ref{eq:mcv}) is
universal. As we have already mentioned, the Brody distribution itself only approximately characterizes the statistics for intermediate values of $w$.  Nevertheless, we can say that our crossover curve shares some general features in common with the curve calculated in~\cite{Jachymski:2015pra} using a QDT approach. This is despite some important differences in the two
calculations. The most important of these is that Reference~\cite{Jachymski:2015pra} assumes that the
closed-channel sector has already been diagonalized and tracks the
transition to chaos by increasing the resonance width (controlled in
our model by $g_{oc}$), while our model holds $g_{oc}$ fixed and
varies $g_{cc}$. Reference~\cite{makrides:fractal_2018} shows a similar crossover curve with increasing magnetic field in lanthanide dimers using \textit{ab initio} methods.

Reference~\cite{Jachymski:2015pra} proposed that the spectral statistics observed in experiments with Erbium and Dysprosium may be more likely to obey the semi-Poisson distribution Eq.~(\ref{eq:semipoisson}). We see from a visual inspection of panel (b) of Fig.~\ref{fig:distributions} that indeed the semi-Poisson curve more closely matches the histogram for this intermediate Brody parameter. A more quantitative measure of the "goodness of fit" is the reduced $\chi^2$. For a histogram with bin counts $h(i)$ for $i = 1,..,n$, the reduced $\chi^2$ is given by \begin{equation}
    \chi_r^2 =\frac{1}{n-p} \sum_i^n{\frac{(h(i)-e(i))^2}{e(i)}}
\end{equation}
where $e(i)$ is the number of bin counts predicted by the proposed distribution and $p$ is the number of fitting parameters. If the data are well-described by the fit, one obtains $\chi^2_r \approx 1$ for a large enough sample;  $\chi^2_r \gg 1$ indicate that the data are unlikely to be distributed according to proposed distribution. In Fig.~\ref{fig:chisqr}, we show $\chi^2_r$ for both the Brody (black circles) and semi-Poisson (red squares) distributions. The value of $w$ that maximizes the log-likelihood function was used for the Brody distribution. The semi-Poisson distribution indeed performs better for intermediate values of the couplings, despite having no fitting parameter, giving a minimum $\chi^2_r$ near $w\approx 0.5$, in rough agreement with Ref.~\cite{Jachymski:2015pra}.  Note however that the semi-Poisson distribution at best gives a $\chi^2_r\approx 4$ which remains large compared to the $\chi_r^2 \lesssim 1$ values achieved by the Brody distribution in the Poisson and GOE limits.

\begin{figure}
\begin{center}
\leavevmode
    \includegraphics[width=0.44\textwidth]{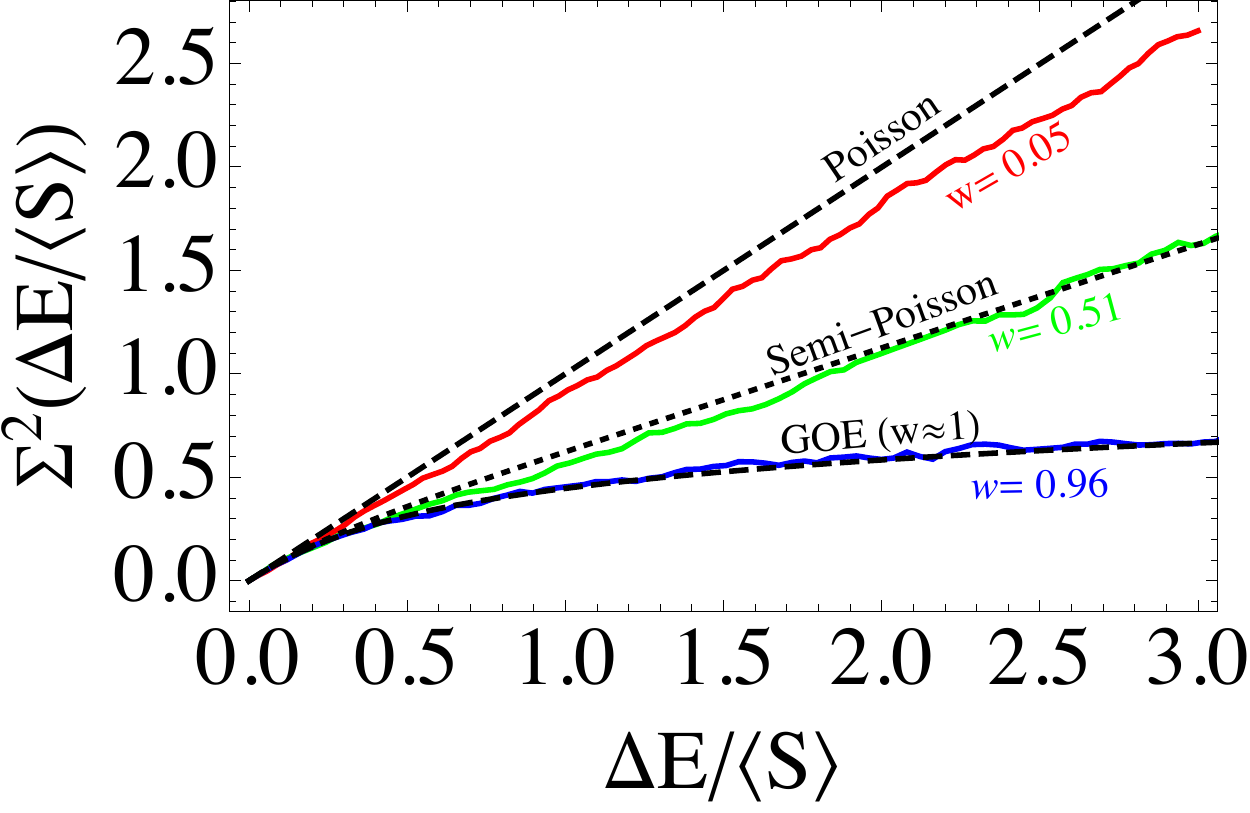}
    \setlength{\unitlength}{\textwidth}
    \end{center}
    \caption{(color online) The variance in the number of resonances within an energy
      window $\Delta E/\langle S\rangle$ is plotted for various
      closed-channel couplings that correspond to $w = 0.05, 0.51,
      0.96$ for the red, green and blue curves, respectively.  The
      dashed curves show the analytical results in the uncorrelated
      limit of Poisson statistics, and the GOE.
      The dotted curve indicates the level number variance for semi-Poisson
      statistics given by Eq.~(\ref{eq:spsigma}). 
    \label{fig:psigma}
    }  
\end{figure}
  
The conjecture of~\cite{Jachymski:2015pra} that perhaps the semi-Poisson distribution is the physical distribution for lanthanide atoms is informed by calculations of the level number variance Eq.~(\ref{eq:sigmadef}), which we discuss in the context of our work now. In Fig.~\ref{fig:psigma}, we show the number variance as a function of the size of the energy window for the same three Brody parameters indicated in the distributions of Fig.~\ref{fig:distributions}. The top black dashed line in Fig.~\ref{fig:psigma} represents the expected number variance for Poisson statistics,
\begin{equation}
\label{eq:sig2poisson}
    \Sigma^2(\epsilon)=\epsilon,
\end{equation}
while the lower dashed black curve is for the GOE. Exact relations for the number variance for various ensembles have been given in~\cite{Brody:1981rmp}. For the GOE, one finds, defining $\epsilon = \Delta E/\langle S \rangle$,
\begin{align}
\Sigma^2_{\text{GOE}}(\epsilon) &= \frac{2}{\pi^2} \left[ \ln{\left(2 \pi \epsilon \right)} + \gamma + 1
     -\cos{\left(2 \pi \epsilon \right)} - \text{Ci}\left(2 \pi\epsilon \right) \right ] \notag \\
      &+ 2\epsilon \left(1-\frac{2}{\pi}\text{Si}\left(2 \pi \epsilon \right)\right) + \left(\frac{\text{Si}\left(\pi \epsilon \right)}{\pi} \right)^2 - \frac{\text{Si}(\pi \epsilon)}{\pi}
\end{align}
where $\text{Si}(z)=\int_0^z{\frac{\sin{t}}{t}dt}$ and $\text{Ci}(z)=\int_{-z}^{\infty}{\frac{\cos{t}}{t}dt}$ are the sine and cosine integral functions, respectively, and $\gamma\approx 0.5772$ is Euler's gamma constant. Finally, the black dotted curve in Fig.~\ref{fig:psigma} shows the expected variance from semi-Poisson statistics,
namely~\cite{GarciaGarcia:2005PRE}: 
\begin{equation}
\label{eq:spsigma}
    \Sigma^2_{\text{SP}}(\epsilon)= \frac{\epsilon}{2} + \frac{1-\exp{(-4 \epsilon)}}{8}
\end{equation}

Our results show that the number variance crosses over from Poisson-like to GOE-like as $g_{cc}$ increases.
Reference~\cite{Jachymski:2015pra} found that as the typical width of the resonances increases to be of order the average level spacing, the number variance curve saturates to Eq.~(\ref{eq:spsigma}).  In our scheme, we hold the width fixed in model units (but see the inset in Fig.~\ref{fig:pbrody}) while increasing the coupling among closed channels only. Unlike Ref.~\cite{Jachymski:2015pra}, our variance curve does not saturate, but smoothly varies from the Poisson limit to the GOE limit. However, our results support Ref.~\cite{Jachymski:2015pra} to the extent that an intermediate Brody parameter of $w\approx 0.5$ gives a number variance most closely matched to Eq.~(\ref{eq:spsigma}). As a point of reference, when $10^4$ points are drawn from the semi-Poisson distribution, the method of maximum likelihood yields a Brody parameter of $w=0.50\pm 0.01$. Binning those $10^4$ points into a histogram with gives a $\chi_r^2\approx 6$ for the Brody distribution, and (not surprisingly) $\chi_r^2\approx 1$ for the semi-Poisson distribution.

\section{Determining the model parameters from physical properties \label{PhysContext}}
Here we discuss how to directly determine our model parameters from properties of a physical system, such as ultracold lanthanide atoms or molecules.  Determining the model parameters requires some experimental (or more microscopic theoretical) input, and  the greater the detail of the experimental data, the more precisely the model may be specified. This is  analogous to the situation with all pseudopotentials used throughout ultracold matter. One common example is the delta function pseudopotential $g \delta(\vec{r})$, where one determines $g$ by matching an experimentally measured scattering length. Another common example is the  two-channel square-well model, in which one can match the scattering near a resonance by matching the experimentally measured resonance position and width~\cite{Chin:2010rmp}. Regardless of the experimental data and pseudopotential used, the key requirement for such models to be useful is that they can predict properties that are not used as input, such as how scattering is affected by external fields, effective lattice model parameters~\cite{Docaj:2016prl,Wall:2017pra,Wall:2017pra2}, or few- and many-body properties. 

We will show that the coupled square-well model suffices to reproduce the low-energy ($kr_0\ll1$) scattering in collisionally complex systems, and give explicit formulas for determining the model parameters, at least in the  non-overlapping (closed-channel) resonance limit. This is expected to be a relevant limit for both lanthanides~\cite{Maier:2015pra,Frisch2014} and molecules~\cite{Mayle:2012pra,Mayle:2013pra}.  The model could also be useful in the overlapping resonance limit, but here simple formulas are not available. Instead, the best approach may be to do a numerical least-squares fit that chooses model parameters to minimize the error between the predicted and observed scattering cross section for all measured energies.

A complete set of low-energy scattering data for the non-overlapping resonance region would be the resonance positions, widths, and background scattering length. We'll refer to the values of these that we wish to match as $\varepsilon^{(i)}$, $\gamma^{(i)}$, and $a_{\text{bg}}$, respectively.  We'll first sketch the idea of how to reproduce these with the square well model, then give the details. The procedure is to take the closed channels as decoupled from each other  $(C_{ij}\rightarrow 0$ where both $i,j \in c$)    and then: (i) For each resonance, choose an appropriate closed-channel depth $D_i$ so that the $E_0^{(i)}$ match the desired resonance positions. (ii) Adjust the open channel depth $D_N$ to match the background scattering length $a_{\text{bg}}$. (iii) For each resonance, adjust the open-closed couplings  $C_{Ni}$ so that the  $\Gamma^{(i)}$ match the desired resonance width. 

The detailed equations to determine these coefficients follow. We focus on $s$-wave scattering for simplicity, but all of the results generalize straightforwardly to other angular momenta.  For non-overlapping resonances, we may determine the parameters for each resonance separately. To further simplify, we take the closed channel thresholds to infinity, $\Delta_i\rightarrow \infty$ for $i\in c$; the model remains flexible enough to reproduce the desired scattering data. The only remaining parameters are the $D_i, D_N$ and the couplings $C_{Ni}$ between the open and the closed channels. Now we carry out the three steps above. 

\textit{(i) Determining the closed-channel depths $D_i$.} For each resonance, we match a 
bound state of a unique closed channel
to the desired resonance position $\varepsilon^{(i)}$.  For finite $\Delta_i$ the bound states must be solved numerically, 
since we are taking $\Delta_i=\infty$,  the channel $i$ potential is a particle in a box, and therefore the bound eigenenergies for channel $i\in c$ are 
$ -D_i + n^2\pi^2 \epsilon_0 $
where $n$ is an integer. 
We match the lowest bound state in each closed channel (associated with $n=1$), to the desired resonance energy $\varepsilon^{(i)}$, giving 
\begin{equation}
D_i = \pi^2 \epsilon_0 - \varepsilon^{(i)},\label{eq:determined-Dc}
\end{equation} 
where $\epsilon_0$ is defined in Eq.~(\ref{eq:eps0}). This introduces infinitely many resonances associated with $n>1$, but these  are at high-energy [the next resonance is 3$\pi^2\epsilon_0$ higher in energy], and hence are negligible for the low energy scattering.

\textit{(ii) Determining the open-channel depth $D_N$.}  For each resonance, we match the background scattering length to the desired value $a_{\text{bg}}$. The background scattering length is observed by measuring the scattering length at energies far from the resonances, and in this regime the open channel decouples from the closed channels. Therefore $a_{\text{bg}}$ is the scattering length obtained from the single particle problem involving the open channel only. This is a finite attractive square well, for which the scattering length is given by
\begin{equation}
a_{\text{bg}} = r_0 \left( 1- \frac{\tan(K_N r_0)}{K_N r_0}\right) \label{eq:determined-Do}
\end{equation}
with $K_N r_0 = \sqrt{D_N/\epsilon_0}$. 
Given $a_{\text{bg}}$ this is a simple transcendental equation with one variable and is easily numerically solved for $a_{\text{bg}}$.

\textit{(iii) Determining the open-closed couplings $C_{Ni}$.} For each resonance, we match the resonance width to  the desired one $\gamma^{(i)}$ by choosing $C_{Ni}$ appropriately.  Since the resonances do not overlap, one can use the single closed-channel formulas from Ref.~\cite{lange:determination_2009}. Specifically, we focus on matching the Feshbach resonance width, the magnetic field scale on which the scattering length changes relative to the background scattering length, but this can be converted to whatever width is conveniently measurable. In this case, the Feshbach width is given by (for $i\in c$)
\begin{equation}
\gamma^{(i)} =   2 \frac{(a_{\text{bg}}-r_0)^2}{\delta \mu \; a_{\text{bg}}r_0}\left(\frac{C_{Ni}}{D_N-D_i} \right)^2 D_i,
\end{equation}  
with $\delta \mu$ the difference in magnetic moments of the open channel $N$ and closed channel $i$. This can be solved to determine $C_{Ni}$:
\begin{equation}
C_{Ni} = \frac{D_N-D_i}{a_{\text{bg}}-r_0} \sqrt{\frac{\gamma^{(i)} \delta \mu \; a_{\text{bg}}r_0}{2D_i }}. \label{eq:determined-Cob}
\end{equation} 
Together, Eqs.~\eqref{eq:determined-Dc}, \eqref{eq:determined-Do} and \eqref{eq:determined-Cob}  determine the multichannel square well parameters from physically measurable scattering data. 

Although the above procedure assumes a complete knowledge of the mentioned scattering data, partial data can still be usefully employed to more qualitatively determine the square model parameters. 
Presently, this is especially important as even the most advanced experiments and theoretical calculations on diatomic polar molecule collisions cannot yet provide the detailed scattering data. If statistical properties of the spectra are known---either from experiment or from \textit{ab initio} calculations, such as \cite{Croft:2017pra} ---then one may use the prescription outlined in Section~\ref{sec:statistical-model} to reproduce the desired target density of states and Brody parameter. 
   One may subsequently obtain as a prediction of the model other statistical properties such as the number variance or spectral rigidity, and use the resulting effective pseudopotential.   

 \section{Conclusions \label{conclusions}}

We have introduced a multi-channel square-well model that is simultaneously simple to use and capable of incorporating the proliferation of resonances necessary to model collisionally complex ultracold matter, such as molecules and lanthanide atoms. It is the simplest finite-ranged effective model for collisions with many resonances.  We have reduced the two-particle scattering solutions of this model to $2N+1$ linear equations with analytic coefficients [Eq.~\eqref{eq:Axb}] and presented a sample of typical results. 

This model builds a framework for researchers to study a variety of physical properties related to scattering physics of systems involving a large density of states for the collision complex. The model is semi-analytic and comparatively easy to use.  It presents advantages over more accurate, but computationally intensive \textit{ab initio} methods for applications where the detailed structure of the spectrum may not be important, but particular statistical properties must be treated correctly. It also provides a useful alternative to zero-range multichannel models~\cite{Docaj:2016prl,Wall:2017pra,Wall:2017pra2}, which neglect finite-range effects and require a tedious regularization associated with working in the zero-range limit. 

We also have also introduced a choice of model parameters $D_i$, $C_{ij}$, and $\Delta_i$, given by Eq.~\eqref{eq:model1}. The couplings $C_{ij}$ are drawn from a Gaussian distribution, while the $\Delta_i$ are chosen from a different, uniform distribution.   Although crude, this choice is intended to realistically capture the statistical properties of systems with complex collisions: Many pairs of channels are coupled with comparable magnitudes $C_{ij}$, while the channel thresholds  $\Delta_i$ are drawn from a uniform distribution with a different energy scale set by the number of channels and the desired density of states.

By solving the scattering properties of the multi-channel square-well model with parameters chosen from the statistical ensemble, we naturally capture  the crossover from integrable to chaotic behavior as a function of the closed-channel coupling relative to the average channel spacing $g_{cc}/\braket{S}$. This is evidenced by fitting the calculated resonance position spacing distribution to the Brody distribution.
We find good fits to these distributions, and show that the Brody parameter evolves from integrable to chaotic as a function of $g_{cc}/\braket{S}$. A natural next step is to calculate the statistical properties of the resonance widths.

Thus we expect this combination of scattering model and ensemble of model parameters to provide a foundation for research on the integrable-chaotic scattering crossover. Specifically, it will allow researchers to explore the consequences of complex collisional interactions without the much more onerous -- and often intractable --  use of a more complex multi-channel collision model.   For example, this model could be used to explore the effect of this integrable-chaotic scattering crossover on Efimov physics or on the many-body phase diagram of molecules.

With some modest extensions, this model can capture other important physical phenomena. For example:
(i) It can be straightforwardly generalized to include the influence of external fields.  By adding   
magnetic or electric dipole moments on both thresholds and depths of the potential,  Eq. (\ref{eq:mcv}) is only slightly modified. 
This would, for example, provide an analog for the magnetic field-tuned resonances in lanthanide collisions where chaotic scattering has been observed \cite{maier:emergence_2015,Frisch2014,Maier:2015pra}. 
(ii) It can be generalized to include many open channels. This requires a slight 
change to Eq.~(\ref{eq:exterior-solns}), where the asymptotic form of the wavefunction becomes:
$\psi_i\rightarrow c_i f_i(k_ir) - \sum s_{ij} g_j(k_jr)$ where $i,j \in o$.
Now there is a separate scattering solution for each possible entrance channel. The coefficients $s_{ij}$ then capture both elastic and inelastic scattering processes.
(iii) Finally, it can be used to study the statistical distribution of resonance widths under various conditions, which appears to be a timely topic of interest in the nuclear physics community~\cite{Koehler:2010PRL,Koehler:2011PRC,Celardo:2011PRL,Fanto:2018PRC}.

\acknowledgments 
N. M. thanks J. F. E. Croft for useful discussions pointing toward Ref.~\cite{Walker:1989JCP}.
N. M. and C. T. acknowledge support in part by the National Science Foundation under Grant No. NSF PHY11-25915. 
C. T. acknowledges support from the Advanced Simulation, and Computing and
LANL, which is operated by LANS, LLC for the NNSA of the
U.S. DOE under Contract No. DE-AC52-06NA25396.
K.R.A.H acknowledges support  from  funds
from the Welch Foundation,  grant no.  C-1872, and he
thanks the Aspen Center for Physics, which is supported
by the National Science Foundation grant PHY-1066293,
for its hospitality while part of this work was performed.  

\bibliography{refs}
\end{document}